\documentclass[aps,pra,showpacs,twocolumn,superscriptaddress,nolongbibliography,floatfix]{revtex4-2}
\usepackage{graphicx}
\usepackage{amsmath}
\usepackage{gensymb}
\usepackage{xparse}
\usepackage{dcolumn}
\usepackage{bm}
\usepackage[colorlinks=true, citecolor=blue,allcolors=blue]{hyperref}
\usepackage{physics}

\begin{document}
\title{Time-resolved 3D momentum spectroscopy in continuous wave atomic photoionization experiments
} 

\author{K.\,L.~Romans}
\affiliation{Physics Department and LAMOR, Missouri University of Science \& Technology, Rolla, MO 65409, USA}

\author{ B.\,P.~Acharya}
\affiliation{Physics Department and LAMOR, Missouri University of Science \& Technology, Rolla, MO 65409, USA}

\author{ A.\,H.\,N.\,C.~De Silva}
\affiliation{Physics Department and LAMOR, Missouri University of Science \& Technology, Rolla, MO 65409, USA}

\author{ K.~Foster}
\affiliation{Physics Department and LAMOR, Missouri University of Science \& Technology, Rolla, MO 65409, USA}

\author{ O.~Russ}
\affiliation{Physics Department and LAMOR, Missouri University of Science \& Technology, Rolla, MO 65409, USA}

\author{D.~Fischer}
\affiliation{Physics Department and LAMOR, Missouri University of Science \& Technology, Rolla, MO 65409, USA}

\date{\today}

\begin{abstract}
    An experimental continuous-wave (cw) pump-probe scheme is demonstrated by investigating the population and photoionization dynamics of an atomic system. Specifically, $^6$Li atoms are initially prepared in optically pumped $2^{2}S_{1/2}$ and $2^{2}P_{3/2}$ states before being excited via multi-photon absorption from a tunable femtosecond laser. The subsequent cascade back to the ground state is analyzed by ionizing the atoms in the field of a cw optical dipole trap laser. Conventional spectroscopic methods such as standard cold-target recoil-ion momentum spectroscopy (COLTRIMS) or velocity map imaging (VMI) cannot provide simultaneous momentum and time-resolved information on an event-by-event basis for the system investigated here. The new approach overcomes this limitation by leveraging electron-recoil ion coincidences, momentum conservation, and the cyclotron motion of the photoelectron in the magnetic spectrometer field. This enables the reconstruction of ionization times and time-of-flight of the charged target fragments with nanosecond resolution. As a result, not only can three-dimensional photoelectron momentum vectors be determined, but the (incoherent) population dynamics of the atomic system also become accessible. Future applications exploring coherent atomic dynamics on the nanosecond timescale would not only expand the scope of time-resolved spectroscopy but can also aid in developing coherent control schemes for precise atomic manipulation.
\end{abstract}


\maketitle

\section{Introduction}
Photoelectron angular and energy spectroscopy provides detailed insights into atomic and molecular structure, electronic correlations, and quantum-mechanical few-particle dynamics (e.g., \cite{Manson1982, Reid2003}). Beyond its historical significance in the development of quantum scattering theory \cite{Bethe1957}, it has---along with advancements in new light sources and pulse shaping techniques---become one of the foremost tools for investigating ultrafast electronic dynamics and advancing coherent control schemes in atomic and molecular systems (e.g., \cite{Liu2024, Maroju2023, Vrakking2021, Silva2021, Eickhoff2021, Pazourek2015}). The experimental study of such processes relies on spectrometers capable of mapping three-dimensional photoelectron momentum distributions over the full solid angle. In the vast majority of experiments, this is achieved using one of two techniques: \emph{cold-target recoil-ion momentum spectroscopy} (COLTRIMS) also known as \emph{Reaction Microscope} or ReMi (e.g., \cite{Doerner2000, Ullrich2003, Moshammer2003, Fischer2019}), or \emph{velocity map imaging} (VMI) \cite{Eppink1997}). These techniques offer high efficiency, short measurement times, and comprehensive data collection.

For both of these methods, the charged target fragments (electrons and recoil ions) are extracted from the reaction volume by an electrostatic field and detected on two-dimensional position-sensitive detectors. A key difference between the two is that COLTRIMS uses the time-of-flight information of the particles and their position on the detector to reconstruct their three-dimensional momentum vectors. This typically requires pulsed photon sources to provide a time reference, and it is incompatible with photoionzation by continuous wave (cw) radiation. In contrast, VMI directly maps the three-dimensional momentum distribution of the particles onto the two-dimensional detector with high resolution. The full three-dimensional information is then reconstructed using tomography and an Abel inversion \cite{Dasch1992, Vrakking2001}. While VMI can be used with cw light sources and provides momentum distributions integrated over many events, it sacrifices information about the ionization time and the complete momentum vector for individual events.

In this paper, we report on a new approach that enables the measurement of three-dimensional photoelectron momentum vectors on an event-by-event basis without relying on an external time reference, such as a pulsed photon source. This technique is based on the coincident detection of the photoelectron and recoiling target ion following a cw photoionization event in a ReMi. In our measurement, we exploit the fact that the total momentum transfer in the photoionization process is negligibly small, meaning the electron and recoil ion momenta sum to zero. The low background in our measurement, along with the excellent imaging properties of our spectrometer, allows us to reconstruct the time information of the ionization event based on the relative positions of the electrons and recoil ions on the detectors.

We demonstrate the capabilities of our new approach by investigating aspects of an ionization scheme that remain inaccessible with both conventional COLTRIMS and VMI. Specifically, we study the photoionization of $^6$Li Rydberg atoms out of an optical dipole trap in a continuous pump-probe scheme. The lithium atoms are first cooled and confined in a near-resonant all-optical trap \cite{Sharma2018}, creating a sample of cold atoms in either the 2$S$ ground state or a polarized 2$P$ state. A femtosecond laser pulse excites the valence electron to states with principal quantum numbers ranging from $n=6$ to $>10$. Throughout this process, the atoms remain exposed to the continuous-wave field of an infrared dipole trap laser. The fragments of atoms ionized in this field are collected using a ReMi \cite{Hubele2015}. A data-analysis algorithm reconstructs the ionization time of atoms in the continuous-wave field with nanosecond resolution. Consequently, this approach grants access not only to photoelectron angular and energy distributions but also to time-dependent population dynamics driven by spontaneous decay and photoionization.

\section{Experimental Setup}
This experiment consists of four key components. First, the target lithium atoms are laser-cooled in an all-optical trap (AOT), a derivative of the standard magneto-optical trap (MOT) \cite{Sharma2018}. Second, a tunable femtosecond (fs) laser excites the atoms to Rydberg states. Third, a near-infrared continuous-wave optical dipole trap (ODT) laser ionizes the excited atoms. Finally, a reaction microscope (ReMi) measures the momentum of ionization fragments over the full solid angle \cite{Hubele2015}. These techniques are described in detail elsewhere; here, we provide only a brief overview of the specifics relevant to the present experiment. 

The AOT consists of three mutually perpendicular pairs of counterpropagating continuous-wave laser beams, slightly red-detuned from the resonance between the 2$^2S_{1/2}$ ground state and the 2$^2P_{3/2}$ excited state. Similar to standard magneto-optical traps, cooling is achieved through Doppler cooling in an optical molasses \cite{Metcalf2007}. However, unlike conventional MOTs, the AOT does not require position-dependent Zeeman shifts in an inhomogeneous magnetic field to trap the atoms. While the exact trapping mechanism is not yet fully understood, it likely arises from a complex interplay of the atomic multi-level structure, interference-induced intensity patterns, and the non-monochromaticity of the laser beams \cite{Sharma2018}. This trap configuration allows for optical pumping, which, in steady state, leaves approximately 25\% of the $\sim10^7$ trapped atoms in the excited and polarized 2$^2P_{3/2}$ state, with their orbital angular momentum aligned along the extraction direction of the ReMi (i.e., $m_\ell=+1$). The remaining 75\% of the atoms populate the 2$^2S_{1/2}$ ground state.

The femtosecond laser system is based on a Ti:Sa oscillator, generating broadband pulses ($\sim 600-1200$\,nm) with durations typically between 5-8\,fs and a pulse energy of $\sim5$\,nJ at a repetition rate of 80\,MHz. A relatively narrow spectral band is amplified through two non-collinear optical parametric amplifier (NOPA) stages. Further details on the system can be found in \cite{Nish20}. After both NOPA stages, the output delivers up to 3\,W of average power at a 200\,kHz repetition rate, with a central wavelength tunable between 660-1000\,nm (set to $740\pm20$\,nm for this experiment) and a pulse duration of approximately 35 fs. In the present experiment, the beam is circularly polarized (co-rotating with the electron current density of the excited atomic 2$^2P_{3/2}$ state in the AOT) and focused to a waist of $\sim 50$\,$\mu$m, with an average power of 150\,mW at the reaction volume.

The ODT laser source is an industrial-grade IPG Photonics YLR-series fiber laser. It is a diode-pumped ytterbium fiber laser with a continuous-wave output, a power range of 20-200\,W, and a near-infrared wavelength of $1070\pm 5$\,nm. While this laser is generally suitable for generating a potential well to confine atoms in an optical dipole trap \cite{Grimm2000}, in the present experiment, it was used solely to ionize atoms previously excited by the femtosecond laser. During the experiment, the laser operated at 100 W and was spatially overlapped with the fs-laser in the reaction volume, with a focal waist of approximately 50\,$\mu$m.

The ReMi used in this experiment is described in detail in \cite{Hubele2015}. Electrons and recoil ions are extracted in opposite directions and guided onto position-sensitive microchannel plate detectors, which are located 430\,mm from the reaction volume and have a diameter of 80\,mm. In most conventional COLTRIMS setups, particles are extracted using an electric field, followed by a field-free drift region that satisfies the Wiley-McLaren time-focusing condition \cite{Wiley1955}. This ensures that, to first order, the measured time-of-flight of the particles is independent of their initial position. In the present experiment, this condition is not required, as time-of-flight information is not used to calculate momentum vectors. Instead, a uniform electrostatic field was applied throughout the entire spectrometer region to minimize fringe field effects and optimize the imaging properties of the setup. In this experiment, the electric field strength was approximately 1\,V/cm. Additionally, a homogeneous magnetic field parallel to the electric field was applied with a strength of 4\,Gauss.

\section{Data Analysis}
The fundamental principle of COLTRIMS is to determine the primordial momentum vectors of charged target fragments ionized within a small reaction volume by analyzing their time-of-flight and detected positions after traversing well-defined electrostatic and magnetic spectrometer fields. The general concept of the present experiment follows a similar approach, with the following data directly recorded for each ionization event:
\begin{itemize}
    \item $(x_\mathrm{d},y_\mathrm{d},t_\mathrm{d})_\mathrm{e}$: The coordinates and arrival time of the electron on the electron detector.
    \item $(x_\mathrm{d},y_\mathrm{d},t_\mathrm{d})_\mathrm{r}$: The coordinates and arrival time of the recoiling Li$^+$ ion on the ion detector.
    \item $t_\mathrm{fs}$: The time of the femtosecond excitation laser pulse, recorded with a photodiode.
\end{itemize}
In conventional COLTRIMS, the photoionization time $t_\mathrm{i}$ is additionally determined by pulsing the ionizing radiation, a crucial step for reconstructing the three-dimensional momentum vectors. However, in the present experiment, the ionization time is not directly accessible and must instead be recovered using the method outlined below.

For the following considerations, we use a Cartesian coordinate system where the reaction volume is at the origin, and the constant electric and magnetic fields, $\vec{E}$ and $\vec{B}$, respectively, are aligned along the $z$-direction. The electron and recoil ion detectors are positioned facing the reaction volume, and they are centered at $(0,0,z_\mathrm{d})$, with $z_\mathrm{d} = \pm 430$\,mm---where the electron detector is located at the negative $z$ position and the ion detector at the positive one. In the presence of the electric and magnetic spectrometer fields, the charged fragments experience the Lorentz force, and their detected positions depend on the primordial momentum $\vec{p}$ as follows \cite{Fischer2019}:
\begin{align}
    x_\mathrm{d}&=\frac{1}{m\omega_c}(p_x\sin(\omega_cT)+p_y(1-\cos(\omega_cT)))\label{eq:xcoord}\\
    y_\mathrm{d}&=\frac{1}{m\omega_c}(p_x(\cos(\omega_cT)-1)+p_y\sin(\omega_cT))\label{eq:ycoord}\\
    z_\mathrm{d}&=\frac{p_z}{m}T+\frac{qE_z}{2m}T^2,\label{eq:zcoord}
\end{align}
where $q$ and $m$ are the charge and mass of the target fragment, respectively, and $\omega_c = \frac{q B_z}{m}$ is its cyclotron frequency in the magnetic field. The time-of-flight $T$ is given by the difference between the photoionization time $t_\mathrm{i}$ and the detection time $t_\mathrm{d}$, i.e., $T = t_\mathrm{d} - t_\mathrm{i}$.

If $T$ (or equivalently, $t_\mathrm{i}$) is known, solving the above equations for the momentum components $p_x$, $p_y$, and $p_z$ for each target fragment is straightforward. In the present case, however, $t_\mathrm{i}$ is unknown. Nevertheless, the momentum can still be determined by solving the system of equations simultaneously for both the electron and the recoil ion, using the additional constraint that the total momentum transfer in a photoionization process is negligibly small:
\begin{equation}
\vec{p}_\mathrm{e}+\vec{p}_\mathrm{r}=0.\label{eq:momsum}
\end{equation}

\begin{figure}[t]
    \centering
    \includegraphics[keepaspectratio, width=0.8\linewidth]{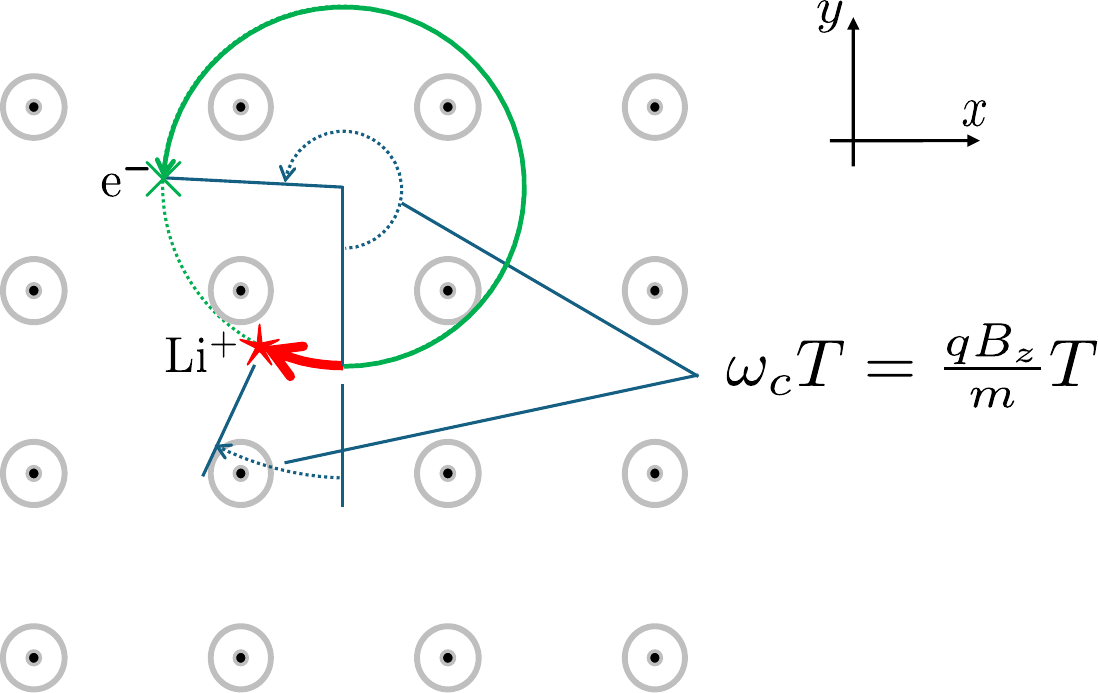} 
    \caption{Example trajectories of an electron and recoil ion in the $xy$-plane for a magnetic field oriented in the $z$-direction (see text).}
    \label{fig:trajectories}
\end{figure}

To better understand this method, we now examine the motion of electrons and recoil ions in the plane perpendicular to the spectrometer fields, i.e., the $xy$-plane. Figure~\ref{fig:trajectories} illustrates example trajectories for electrons and recoil ions with equal momentum magnitudes but opposite directions—specifically, along the positive and negative $x$-axes, respectively.

According to Eqs.~\ref{eq:xcoord} and \ref{eq:ycoord}, both singly charged particles follow circular trajectories with an identical radius of $r=\sqrt{p_x^2+p_y^2}/|q|B_z$, but they rotate in opposite directions (see Fig.~\ref{fig:trajectories}). However, due to their significantly different masses---the $^6$Li$^+$ ion being approximately 11,000 times more massive than the electron---the rotational angles $\omega_cT$ covered by the two particles during their passage through the spectrometer differ substantially. For zero longitudinal momentum ($p_z=0$), this angle follows the proportionality $\omega_cT \propto 1/\sqrt{m}$, which implies that the angle covered by the electron is typically more than 100 times larger than that of the recoil ion. For the electric and magnetic field strengths typically used in COLTRIMS experiments, the electron undergoes several full cyclotron revolutions before reaching the detector, whereas the recoiling target ion rotates much more slowly, covering only a small fraction of a full $2\pi$ rotation. Consequently, the relative positions of the electrons and recoil ions provide information about the cyclotron phase angle of the electron and, therefore, its time of flight $T_\mathrm{e}$.

However, this method can yield ambiguous results, as the equations above are also satisfied when adding $2\pi$ to the electron phase angle (i.e., shifting the ionization time $t_i$ by $2\pi/\omega_c$). To eliminate this ambiguity, the spectrometer fields must be chosen such that all electrons within the relevant photoelectron energy range complete the same number of full cyclotron revolutions---plus an additional variable angle smaller than $2\pi$---before reaching the detector. In the present experiment, the highest observed photoelectron energy is approximately $E_e \approx 1.5$,eV. The electric and magnetic spectrometer fields are set to $E = 1$\,V/cm and $B = 4$\,Gauss. Under these conditions, the electrons complete between two and three cyclotron revolutions before reaching the detector. In contrast, the $^6$Li$^+$ ions cover only a much smaller cyclotron angle of about 8.5°.

In practice, the ionization time $t_\mathrm{i}$ is determined individually for each event using a numerical optimization algorithm. This process involves selecting an initial value for $t_\mathrm{i}$ and iteratively adjusting it to best satisfy Eq.~\ref{eq:momsum}. Specifically, the algorithm minimizes the transverse momentum transfer, defined as $q_\perp=\sqrt{(p_{x,\mathrm{e}} - p_{x,\mathrm{r}})^2 + (p_{y,\mathrm{e}} - p_{y,\mathrm{r}})^2}$ which is derived from  Eqs.~\ref{eq:xcoord} and \ref{eq:ycoord}. The initial seed value for $t_\mathrm{i}$ is chosen based on the assumption that the electron undergoes exactly two cyclotron revolutions, corresponding to a time-of-flight of $T = 4\pi/\omega_c$.

\begin{figure}[t]
    \centering
    \includegraphics[keepaspectratio, width=1\linewidth]{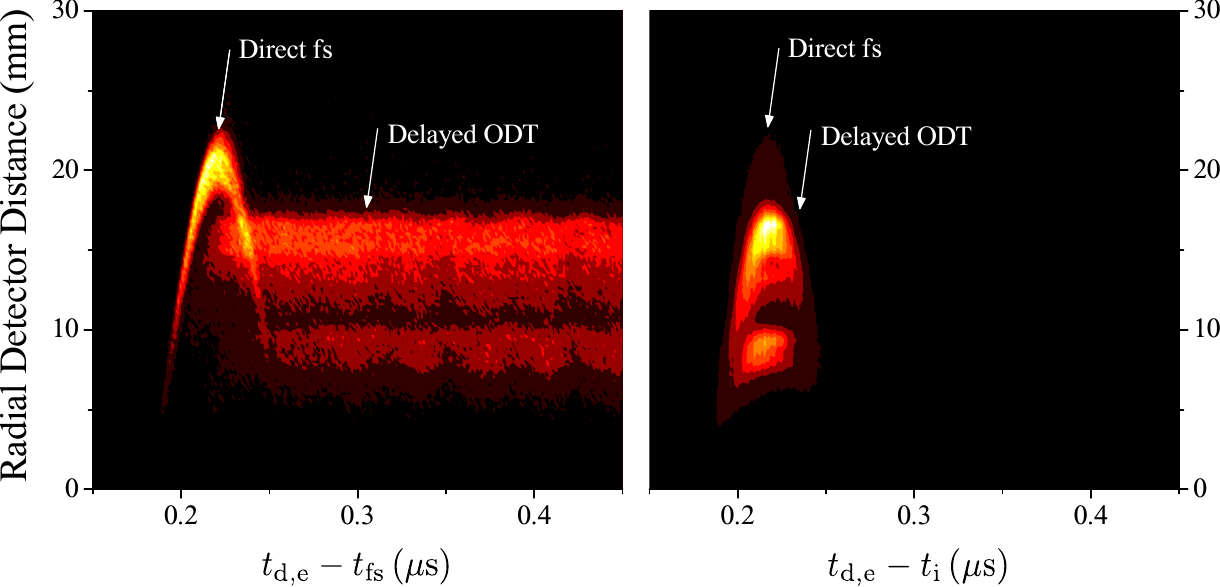} 
    \caption{Electron radial detector distance vs measured time of arrival (left) and derived time of flight (right). The z-axis of the plot corresponds the number of events on a linear scale.}
    \label{fig:radial_vs_time}
\end{figure}

Figure~\ref{fig:radial_vs_time} presents data from the current experiment. In the left panel, electron counts are plotted as a function of the delay between the electrons' arrival at the detector $t_\mathrm{d,e}$ and the femtosecond laser pulse $t_\mathrm{fs}$, as well as the distance of their impact position from the detector center. A well-defined arch is visible between approximately 0.2 and 0.3\,$\mu$s. This structure corresponds to events where the target is directly ionized by the femtosecond laser pulse, producing photoelectrons with energies of roughly 1.3 to 1.5,eV. For these events, $t_\mathrm{fs}=t_\mathrm{i}$, meaning the horizontal axis directly represents the electron time-of-flight $T_\mathrm{e}$. The observed dependence of the radial distance on $T_\mathrm{e}$ results from the electron motion in the spectrometer field, as described by Eqs.~\ref{eq:xcoord} to \ref{eq:zcoord}.

Additionally, the graph exhibits trailing horizontal bands, which originate from the ionization of excited atoms in the continuous-wave (cw) field of the ODT laser. For these events, the time represented on the horizontal axis does not correspond directly to the electron's time-of-flight $T_\mathrm{e}$. This discrepancy arises because there is a variable delay---unique to each event---between the arrival of the femtosecond pulse at time $t_\mathrm{fs}$ and the actual photoionization time $t_\mathrm{i}$. Consequently, no clear correlation exists between the detector radius and the time shown on the horizontal axis.

Applying the previously discussed algorithm to reconstruct the ionization time $t_\mathrm{i}$ allows for the recovery of the electron time-position correlation in the delayed ionization events. The right graph in Fig.~\ref{fig:radial_vs_time} presents the corresponding histogram, where the horizontal axis represents the reconstructed electron time-of-flight, defined as $T_\mathrm{e}=t_\mathrm{d,e}-t_\mathrm{i}$. In this plot, three distinct, stacked arches are visible, each corresponding to a different photoelectron energy band. The outermost arch represents direct femtosecond ionization, while the two inner arches correspond to delayed ionization induced by the ODT laser.    

\begin{figure}[htbp]
\centering
\includegraphics[keepaspectratio, width=0.8\linewidth]{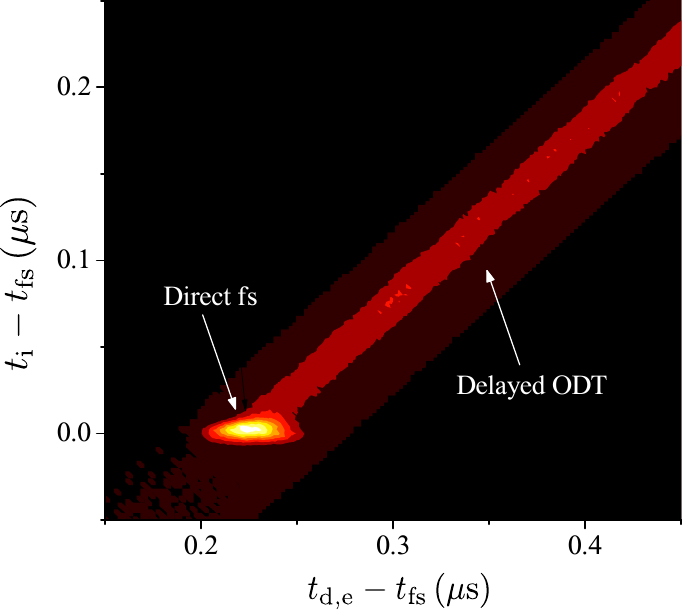} 
\caption{Derived time of ionization vs measured time of arrival with respect to photodiode. The z-axis of the plot corresponds the number of events on a linear scale.}
\label{fig:derive_vs_measured_time}
\end{figure}

The resolution of the time recovery algorithm is fundamentally limited by the precision of the position measurement. Since the algorithm reconstructs the cyclotron phase angle traversed by the electron in the spectrometer fields, its accuracy is expected to be highest for events with large radial positions on both the electron and recoil ion detectors. In other words, the algorithm performs best for electrons with large transverse momenta and cyclotron phase angles $\omega_cT_\mathrm{e}$ that are close to odd integer multiples of $\pi$. 

The time resolution can be evaluated by comparing the reconstructed ionization time to the directly measured ionization time in cases where this is feasible, such as direct femtosecond ionization. This comparison is illustrated in Fig.~\ref{fig:derive_vs_measured_time}, where the electron counts are plotted as a function of the ionization time delay $t_\mathrm{i}-t_\mathrm{fs}$ on the vertical axis and the arrival time at the detector relative to the femtosecond laser pulse $t_\mathrm{d,e}-t_\mathrm{fs}$ on the horizontal axis. For direct femtosecond ionization, the femtosecond laser pulse defines the ionization time, i.e., $t_\mathrm{fs}=t_\mathrm{i}$, causing the counts to accumulate along a narrow horizontal line. The vertical width of this line, which essentially represents the achievable resolution, is approximately $\pm 5$\,ns. This time resolution enables the reconstruction of the electrons' transverse momentum vectors with radial and angular resolutions of about 0.02\,a.u.\ and $\pm 10$\degree, respectively. The tail observed in the graph originates from delayed ODT ionization and has a slope of unity.

\section{Results and Discussion}
\begin{figure}[t]
\centering
\includegraphics[keepaspectratio, width=1\linewidth]{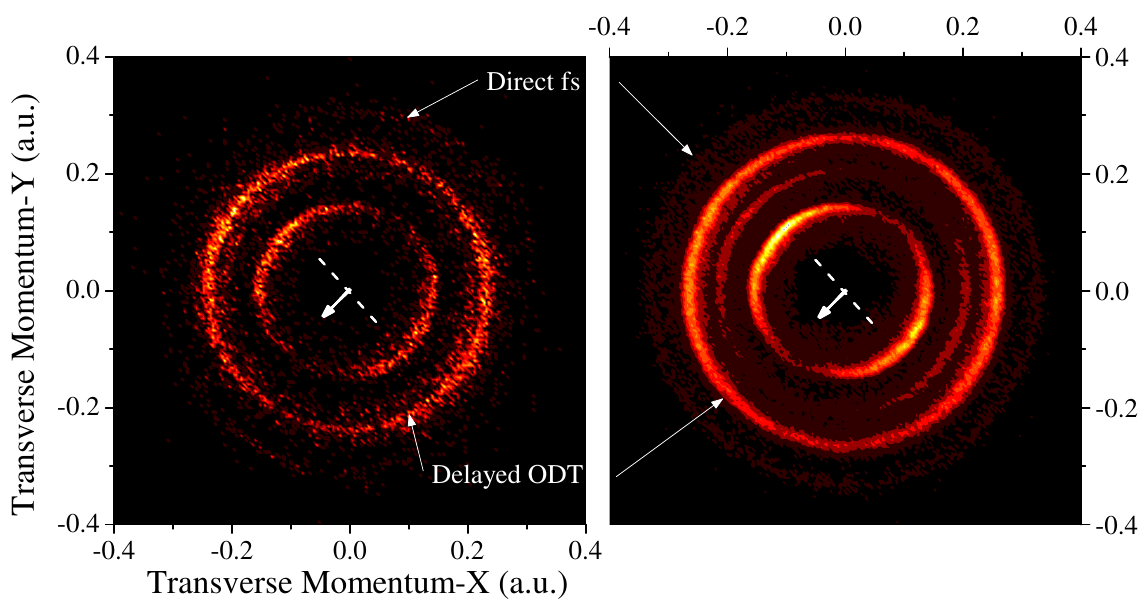} 
    \caption{Electron transverse momentum distributions using derived times from the $2S$ (left) and $2P$ (right) initial states, in atomic units. ODT propagation direction (arrow) and linear polarization axis (dashes) are denoted at each center. The fs-pulse is propagating along the axis into the page. The z-axis of the plot corresponds the number of events on a linear scale.}
\label{fig:transverse_momentum}
\end{figure}

Now that the time of flight can be recovered for any event, the ionization process can be further illuminated. As seen in Fig.~\ref{fig:transverse_momentum}, the transverse momentum is shown for  both the $2^{2}S_{1/2}$ and $2^{2}P_{3/2}$ initial states (hereafter referred to as $2S$ and $2P$). The fs pulse is circularly polarized with the photon spin pointing in the same direction as the orbital angular momentum of the $2P$ electron. The ODT laser is linearly polarized with an electric field direction indicated by the dashed line in the graphs. The radius of each ring corresponds to a given photoelectron energy and the distribution of the ring about the xy-plane conveys information on the final angular state(s). 

\begin{figure}[htbp]
\centering
\includegraphics[keepaspectratio, width=0.8\linewidth]{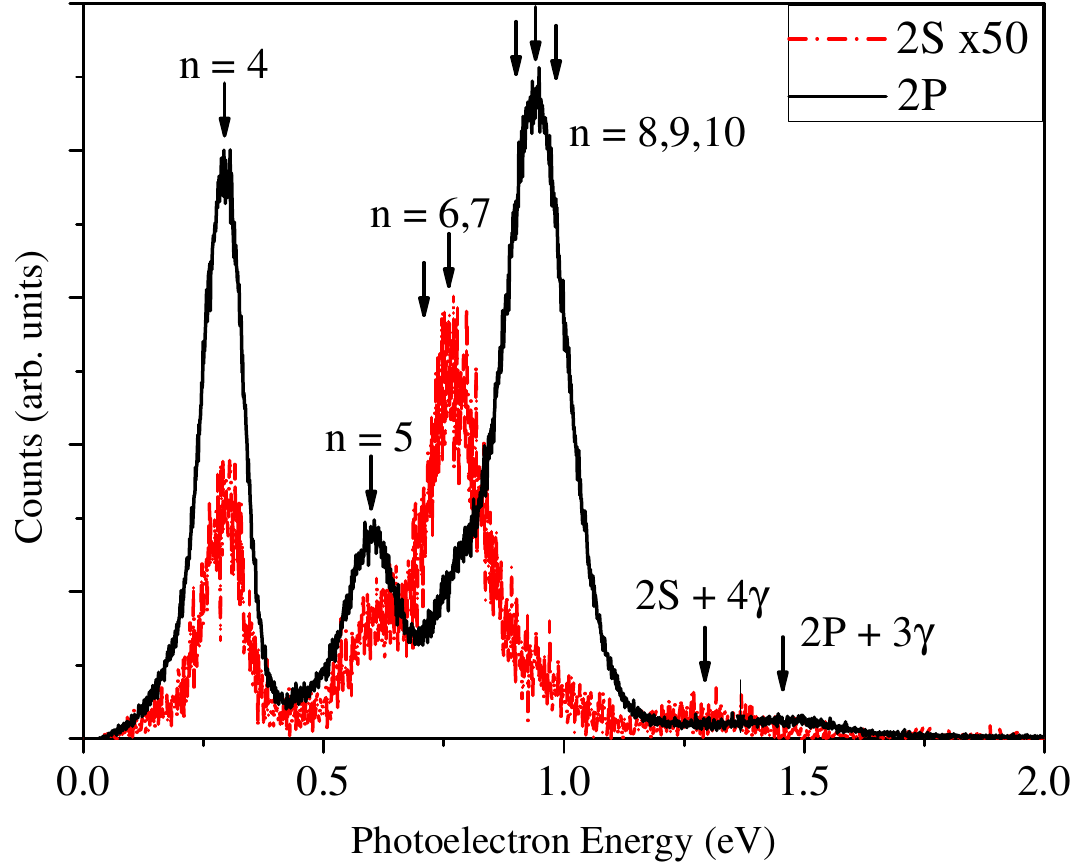} 
\caption{Photo-electron energy distributions for the 2$S$ (red line) and the 2$P$ (black line) initial states. From right-to-left, the bumps are direct fs-ionization with either three or four photons ($\gamma$), and the remaining peaks are the delayed ionization. Estimated principle quantum numbers are denoted for delayed peaks.}
\label{fig:photoelectron_energySpect}
\end{figure}

In Fig.~\ref{fig:photoelectron_energySpect}, detected counts are plotted against the photo-electron energy for both the initial states. Going from right-to-left, the small bumps at about 1.3 and 1.5\,eV correspond to the direct multi-photon ionization from the fs pulse using three or four photons ($\gamma$), respectively.  Each peak thereafter is tied to a ring in the momentum spectra caused by the delayed ionization of the target in the field of the ODT laser. The principle quantum number $n$ of the populated state before the absorption of the ionizing photon is indicated on top of the curves, assuming two or three fs-photons to first excite from the initial states and then a single ODT-photon to ionize.
	 
\subsection{Angular distribution}
\begin{figure}[!htbp]
    \centering
    \includegraphics[keepaspectratio, width=.70\linewidth]{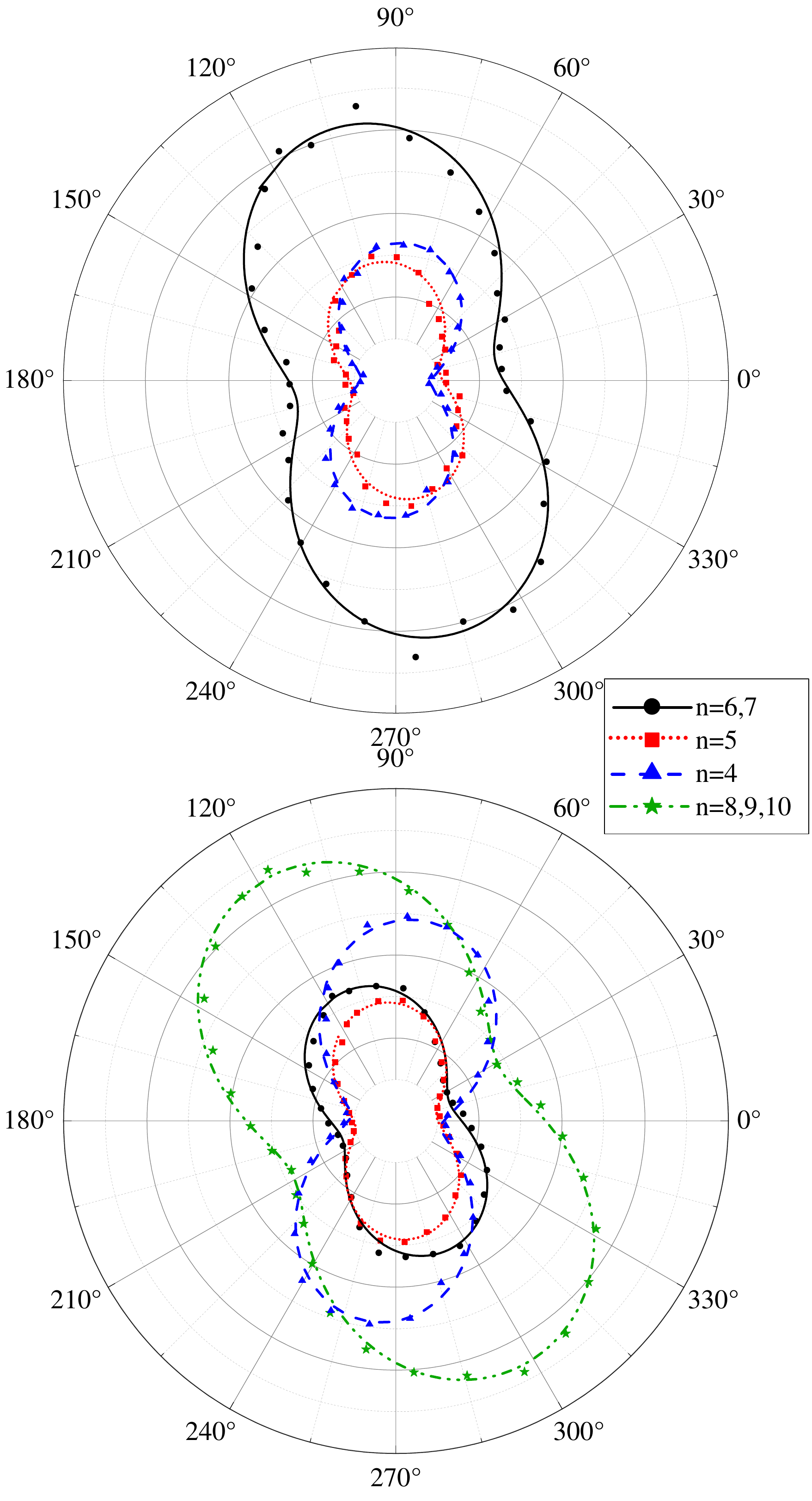}
    \caption{Photo-electron angular distributions of the delayed ionization peaks from the 2S (top) and 2P (bottom) initial states. ODT propagates radially inward along 0$^{o}$ line with polarization along 90$^{o}$ axis, in the transverse xy-plane. Quantization axis is along longitudinal z-axis into the page. Note that data points have been symmeterized.}
    \label{fig:polarplot}
\end{figure}

Information on the final angular momenutm states is encoded in the photo-electron angular distributions (PAD). The corresponding spectra in the $xy$-plane for for delayed ionization channels are shown in Fig.~\ref{fig:polarplot}. For these plots, the ODT laser polarization is along the vertical axis and the quantization axis is chosen perpendicular to the plane of the plot (i.e., the $z$-direction), i.e., the electron polar angle $\theta$ is fixed to 90\degree.

It is immediately clear that the PADs are rotated by an angle about the quantization axis, that their major axes are misaligned to the polarization direction. This deviation can be explained with magnetic dichroism \cite{Acharya2021}. Given that the system begins optically pumped and is excited by a co-rotating laser, it is convenient to move from the ($F$, $m_{F}$) framework of hyperfine splitting into the ($\ell$, $m$) framework of the valence electron. Now, the analysis employed in \cite{Acharya2021} can be directly utilized. 

Each $n\ell$ state of interest can be ionized by the linearly polarized ODT laser into several final magnetic sub-levels, corresponding to both the quantum numbers, $\ell$ and $m$, changing by $\pm$1. The the dependence of the cross section on the electron azimuthal angle $\phi$ can then be expressed in terms of a partial wave expansion \cite{Acharya2021}
\begin{equation}
\begin{split}    
|\Psi_{f}|^2 &= |\sum_{m} c_{m}e^{i m \phi}|^2 \\
&=(\sum_{m} c_{m}\, e^{i m \phi})(\sum_{m'} c_{m'}^{*}e^{-im'\phi})\\ 
&=\sum_{m} (C_{m})^2\\
&+\sum_{m<m'} 2C_{m}C_{m'}\cos[(m-m')\phi + \varphi_{m,m'}],
\end{split}
\label{eq:pad}
\end{equation}
where $c_{m}$ are the complex partial wave amplitudes and the sum is taken over the allowed magnetic sublevels. The last equality is obtained by expressing the complex amplitudes in polar form, $c_{m} = C_{m}e^{i\varphi_{m}}$, using the real parameters $C_{m}$ and $\varphi_{m}$. A clear $\phi$ dependent term arises in Eq.~(\ref{eq:pad}) due to the interference of the paths with differing $m$ and this determines the number of maxima in the distribution. An additional angular shift in the distribution appears due to the relative phase difference between the state amplitudes, $\varphi_{m,m'} = \varphi_{m}-\varphi_{m'}$, which characterizes the phenomenon of magnetic dichroism. This term rotates the whole distribution in the transverse plane about the quantization axis.

In this study, assuming maximal alignment is held, each state is ionized by a single ODT photon corresponding to $(m- m')\phi = 2\phi$ and two terms in the sum. Each state can then be fit to the following simplified form,
\begin{equation}
    |\Psi_{f}|^{2}_\mathrm{fit} = (C_{1})^{2} + (C_{2})^{2} + 2C_{1}C_{2}\cos[2\phi+C_{12}],
\end{equation}
where $(C_{1},C_{2},C_{12})$ are real-valued fitting parameters. These fits are also shown in Fig.~\ref{fig:polarplot} as the various lines and exhibit excellent agreement with the data. 

In principle, one could find the ratio of $C_{1}$ to $C_{2}$ from the ratio of the lengths associated with the major and minor axes for each distribution; this ratio relates the absolute values of the complex amplitudes. The shift of the major axis of the distribution from the vertical axis (polarization) is $C_{12}$ and yields a direct measure of the magnetic dichroism. It must be cautioned that the assumption of a single state per ionization peak quickly breaks down (apparent in Fig.~\ref{fig:photoelectron_energySpect}) as the highest energy peak is a superposition of closely spaced Rydberg states and the lowest is an admixture of states from the resulting cascade. From the present data, it is not easily possible to isolate the contribution from each state. 

\subsection{Time dependent population dynamics}
This experiment enables the study of the system's time evolution using a pump-probe scheme. The system is first excited by a femtosecond laser pulse and then probed via ionization in the cw field of the ODT laser. Here, we demonstrate this capability by investigating the population dynamics of the excited system, which is governed by spontaneous decay and photoionization of the excited states. 

\begin{figure}[htbp]
    \centering
    \includegraphics[keepaspectratio, width=0.8\linewidth]{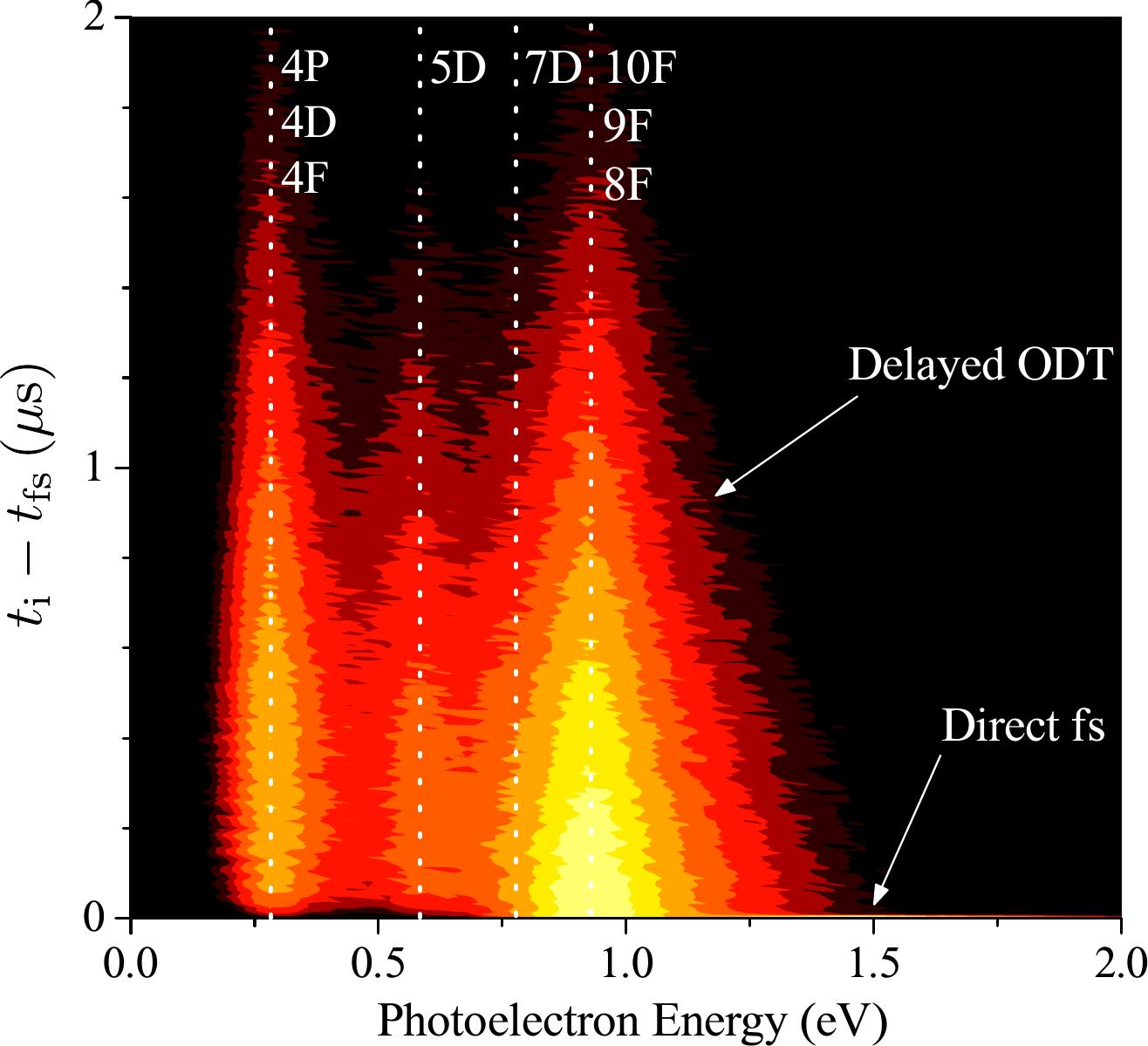}
    \caption{Derived time of ionization vs. photoelectron energy. Centers of delayed energy peaks denoted by dashed liens. The z-axis corresponds to the ionization rate and is on a logarithmic scale.}
    \label{fig:timeVenergy}
\end{figure}

Ignoring direct fs-ionization, there are three and four energy peaks for the 2$S$ and 2$P$ initial states, respectively (see Fig.\ref{fig:photoelectron_energySpect}). The co-rotating fs-photons can only drive the optically pumped atoms into other maximally aligned states (i.e., $\ket{\ell,m}=\ket{\ell,\ell}$). In both the 2$S$ and 2$P$ cases, the absorption of three and two fs-photons, respectively, excites the electron to an $F$-orbital ($\ell=3$), which then decays over time. This time dependence is evident for the 2$P$ set in Fig.\ref{fig:timeVenergy}, where a set of possible participating states is highlighted.

\begin{figure}[htbp]
\centering
\includegraphics[keepaspectratio, width=0.8\linewidth]{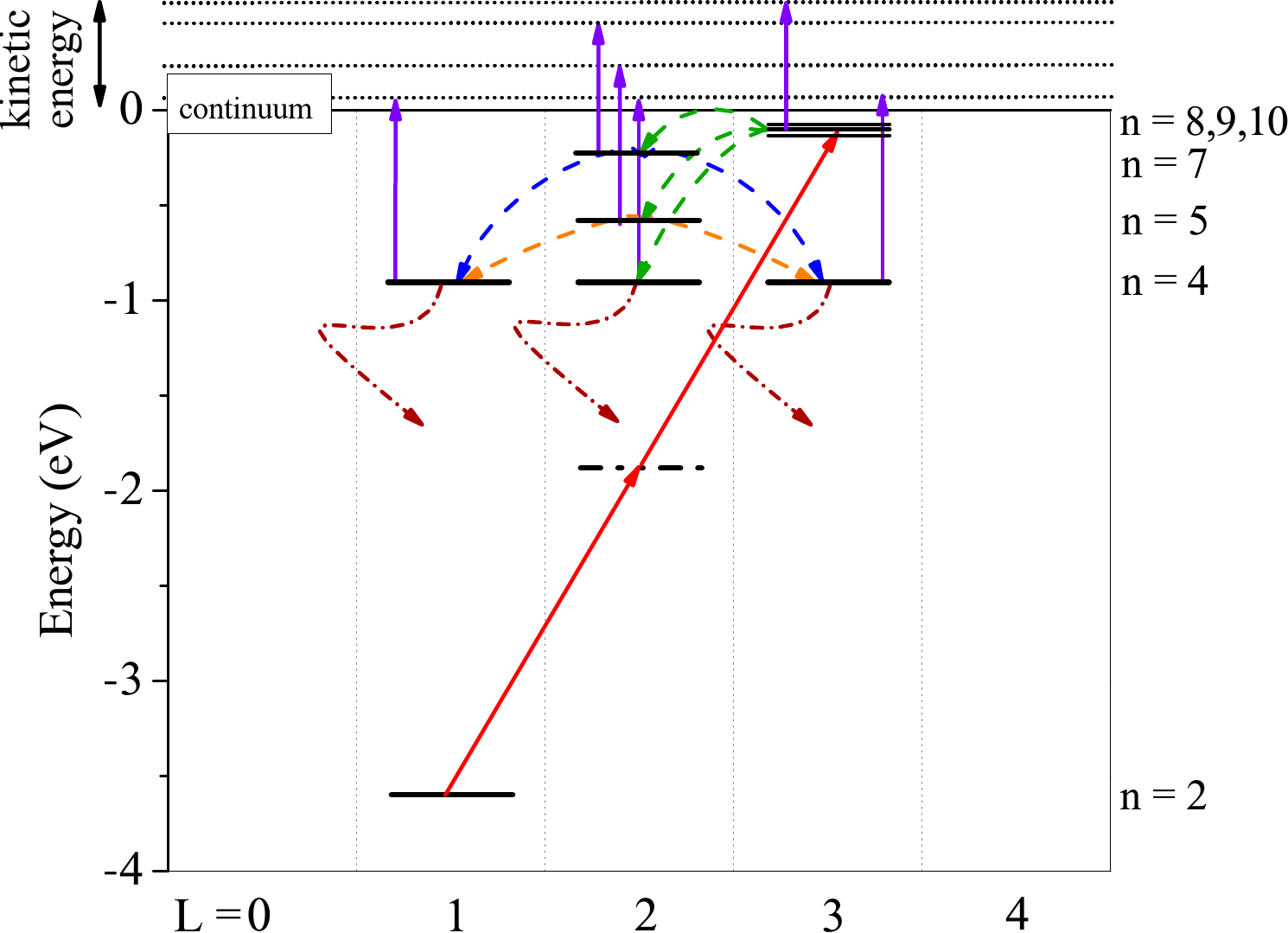} 
\caption{Principle path for $2P$ initial state. Horizontal solid/dashed black lines represent real/virtual atomic states. Other solid lines in color are transitions by the fs (diagonal, red) or ODT (vertical, purple) lasers. Curved dashed lines in color are the various spontaneous decays. Energy peaks seen in Fig.~\ref{fig:photoelectron_energySpect} are the dotted lines above the continuum threshold.}
\label{fig:2P_path}
\end{figure}

Consider first the 2$P$ initial state. The fs-pulse delivers two photons to the atom, exciting it into an $nF$-orbital. Given the spectral width of the fs-pulse and the proximity of the excited states in energy, multiple $n$-states are expected to be populated. From the first peak in the 2$P$ data of Fig.~\ref{fig:photoelectron_energySpect}, the $n=8,9,10$ $F$-states contribute some fraction, $f$, to the total excited population (i.e., $f_{8F}+f_{9F}+f_{10F}=1$). Continuing from right to left, the second peak (visible as a shoulder to the first) is associated with the 7$D$ state, the third peak corresponds to the 5$D$ state, and the fourth peak is attributed to an admixture of the 4$P$, 4$D$, and 4$F$ states. The principal pathways are illustrated in Fig.~\ref{fig:2P_path}.

\begin{figure}[htbp]
\centering
\includegraphics[keepaspectratio, width=0.8\linewidth]{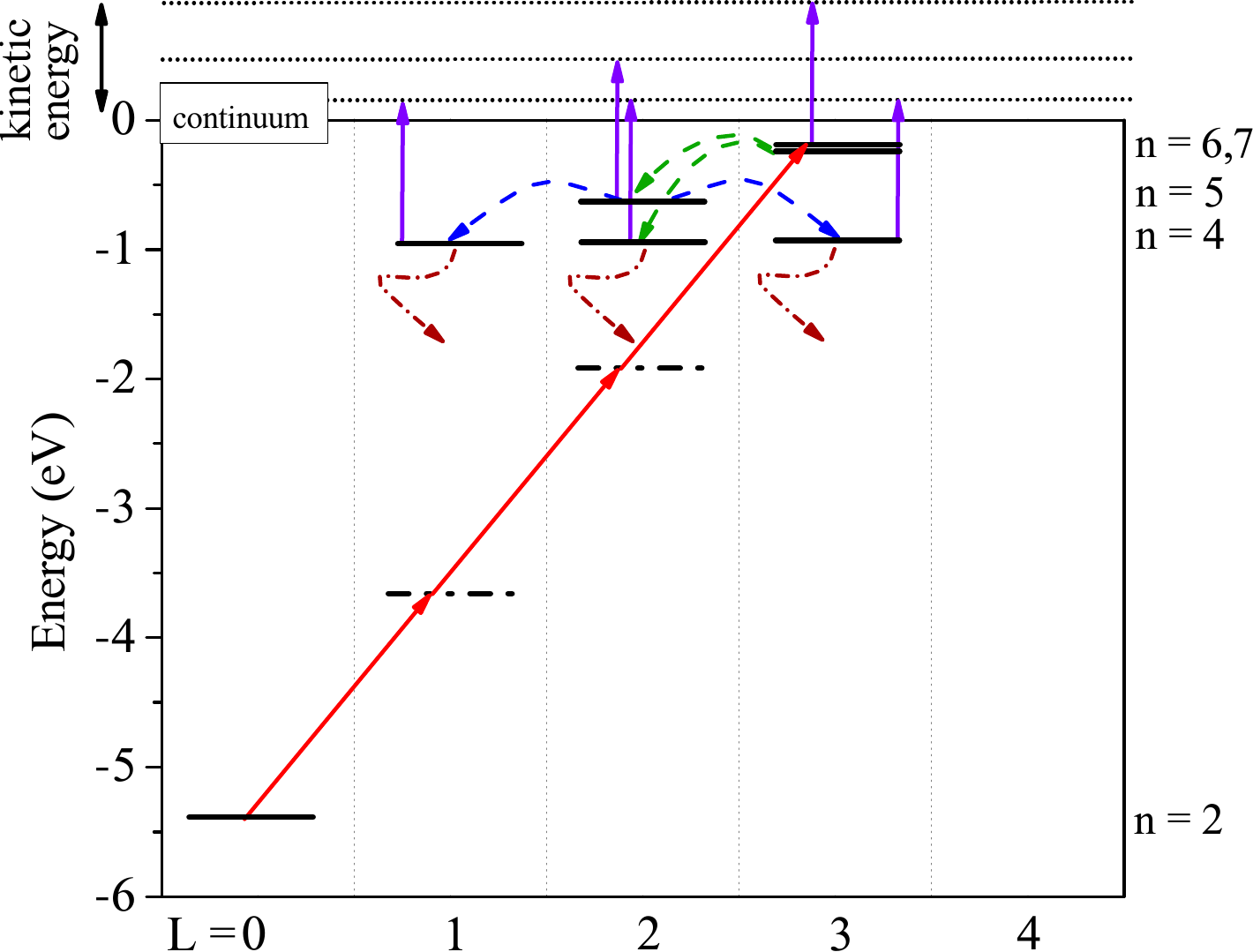} 
\caption{Principle path for $2S$ initial state. The same scheme dontes the states, decays, and transitions as Fig.~\ref{fig:2P_path}.}
\label{fig:2S_path}
\end{figure}

For the $2S$ initial case the system is excited to a mix of the n $= 6,7$ $F$-orbitals via three fs-photons which constitutes the first peak in the red dash-dotted curve in Fig.~\ref{fig:photoelectron_energySpect}. The second peak is tied to the $5D$ state, and the third peak is the same admixture of $n = 4$ states as in the $2P$ set. The principle path for $2S$ is shown in Fig.~\ref{fig:2S_path}. Comparing both paths helps to demonstrate why the photoelectron energy spectra share significant overlap between the $2S$ and $2P$ data sets; nearly the same intermediate states are passed through as the atoms cascade back down towards the ground state. 

The time-dependent population dynamics is described by a set of basic rate equations. Each state in the system, denoted by the index $i$i, has a population of atoms, $N_i(t)$, whose rate of change is governed by the following equation: 
\begin{equation}
    \dv{t} N_{i}(t) = \sum_{j > i} \Gamma_{i,j} N_{j}(t) - (\Gamma_{i} + R_{i}) N_{i}(t),
    \label{eq:population}
\end{equation}
where the $\Gamma_{i,j}$ are the coupling constants that govern the decay of a higher-energy state $j$ into state $i$ via a spontaneous electric dipole transition. The last term accounts for the total depopulation of the $i$-th state due to its spontaneous decay rate $\Gamma_i$ and its photoionization rate $R_i$. For a system of $\mathcal{N}$ interconnected states, there are $\mathcal{N}$ coupled first-order equations of the form given in Eq.~(\ref{eq:population}), which describe the time evolution of the system from a given initial population distribution. The overall system dynamics can then be expressed as:
\begin{equation}
    \dv{t} \boldsymbol{N}(t) = \underline{\boldsymbol{G}} \cdot \boldsymbol{N}(t)
    \label{eq:vector_rate_equation},
\end{equation}
where 
\begin{equation}
    \boldsymbol{N}(t) = \mqty(N_{1}(t) \\ \vdots \\ N_{i}(t) \\ \vdots),
\end{equation}
and 
\begin{equation}
   \underline{\boldsymbol{G}} = \mqty(-(\Gamma_{1} + R_{1}) & \ldots & 0 & \ldots\\  \vdots & \ddots & \ldots & \ldots \\
    \Gamma_{i,1} & \ldots & -(\Gamma_{i} + R_{i}) & \ldots \\ \vdots & \ldots & \ldots & \ddots ). 
\end{equation}
Note that the states are ordered by energy with the highest excited state given the $i = 1$ index. Therefore, all elements above the main diagonal are zero. 

During the decay process, the system is continuously probed by the ODT laser, and ionization rates are measured rather than the population of states directly. To account for this in the model, Eq.~(\ref{eq:vector_rate_equation}) must be modified. A rate matrix $\underline{\boldsymbol{R}}$ is constructed by assigning the photoionization rates of each state to the main diagonal:
\begin{equation}
  \underline{\boldsymbol{R}} = \mqty(R_{1} & \ldots & 0 & \ldots  \\
    \vdots &  \ddots & \ldots & \ldots   \\
    0 & \ldots & R_{i} & \ldots   \\
    \vdots & \ldots & \ldots & \ddots  ),
    \label{eq:rate_Mat}
\end{equation}
such that the ionization rate vector is given by $\boldsymbol{I}(t) = \underline{\boldsymbol{R}}\cdot\boldsymbol{N}(t)$. By multiplying both sides of Eq.~(\ref{eq:vector_rate_equation}) by $\underline{\boldsymbol{R}}$ it follows that the ionization rate equation takes the following form:
\begin{equation}
\begin{split}
    \dv{t} \boldsymbol{I}(t) &= (\underline{\boldsymbol{R}}\cdot \underline{\boldsymbol{G}})\cdot\boldsymbol{N}(t)\\    
   &= \underline{\boldsymbol{G}}\cdot\boldsymbol{I}(t) + \comm\Big{\underline{\boldsymbol{R}}}{\underline{\boldsymbol{G}}}\cdot\boldsymbol{N}(t).
    \label{eq:ionization_rate}
\end{split}    
\end{equation}
It is evident that Eq.~(\ref{eq:vector_rate_equation}) and Eq.~(\ref{eq:ionization_rate}) share nearly identical forms. However, the ionization rate equation includes an additional term involving the commutator of the rate and interaction matrices. This extra term is directly related to the population and contains nonzero elements only below the main diagonal. Therefore, if $\underline{\boldsymbol{G}}$ and $\underline{\boldsymbol{R}}$ are known, Eq.~(\ref{eq:ionization_rate}) can be solved row by row, where the ionization rates $I_i(t)$ depend on $N_j(t)$ values determined in previous rows (i.e., $j<i$).

To describe our experimental data using Eq.~(\ref{eq:ionization_rate}), the coupling constants $\Gamma_{i,j}$, decay rates $\Gamma_{i}$, and photoionization rates $R_i$ must be calculated. We obtained the $\Gamma$ values for the relevant states using the open-source Python package Alkali Rydberg Calculator \cite{Sibalic2017}, while neglecting the dressing of atomic energy levels due to the presence of the oscillating fs and ODT fields. The ionization rates $R_{i}$  were calculated within the framework of a central-potential model, assuming each atom is ionized from a single $n\ell$-state by a single photon.  This calculation begins with the following equation for the ionization cross-section, as found in \cite{Starace1982}:
\begin{equation}
\begin{split}  
\sigma_{n \ell} &= \frac{4 \pi^2 \alpha a_{0}^2}{3} \:\frac{h\,\nu}{2\,\ell+1}\\
&\quad \cross\left[\,\ell\,|\mathcal{R}_{\ell,\ell-1}(\epsilon)|^2 + (\ell+1)\,|\mathcal{R}_{\ell,\ell+1}(\epsilon)|^2\,\right],
\end{split}
\end{equation}
where $\alpha$ is the fine structure constant, $a_{0}$ is the Bohr radius, $h\nu$ is the ionizing photon's energy, and $\epsilon$ the photoelectron's continuum energy with the energy scale. 

The radial overlap matrix elements are given by, 
\begin{equation}
    \mathcal{R}_{\ell,\ell\pm1} = \int_{0}^{\infty} \dd{r}\, U_{n,\ell}(r)\cdot r \cdot U_{\epsilon,\ell\pm1}(r),
\end{equation}
where each reduced radial function $U(r)$ is defined as $r$ times the full radial function of the initial discrete and final continuum state wave functions, respectively. These wave functions were obtained by numerically solving the Schr{\''o}dinger equation using the Lithium potential from \cite{Marinescu1994}, which accounts for core polarization and the quantum defect experienced by the outer valence electron. The numerical solution was carried out using a fifth-order Dormand-Prince (RKDP) method implemented in Mathematica 12.0. Finally, the photoionization rates for the $i^\mathrm{th}$ state were determined by multiplying the cross-section by the incoming ODT photon flux. 
\begin{equation}
    R_{i} = \Phi_{ODT}\cdot\sigma_{n_{i}l_{i}}.
\end{equation}

\begin{figure}[t]
\centering
\includegraphics[keepaspectratio, width=0.9\linewidth]{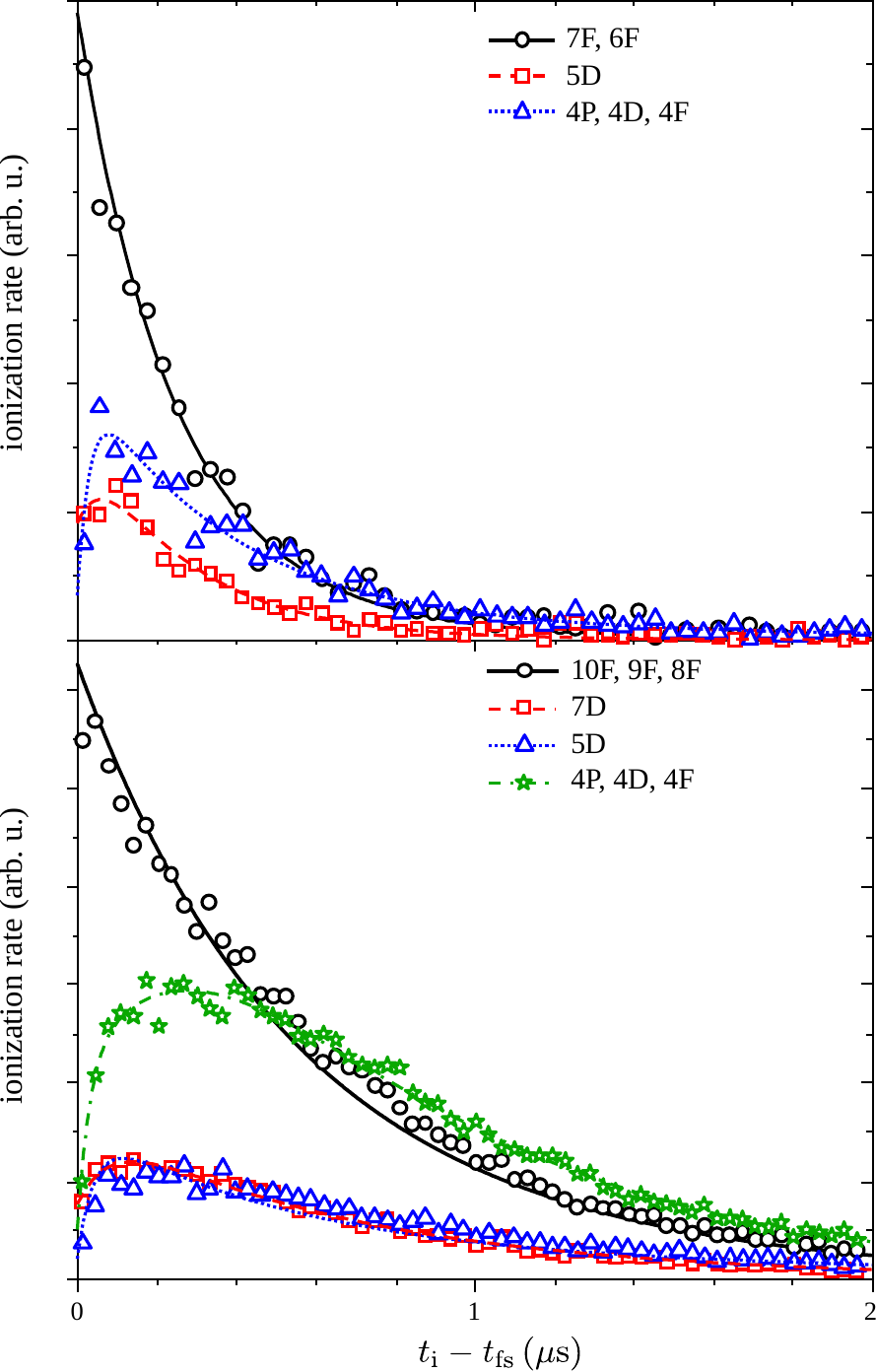}    \caption{Ionization rate versus derived time of ionization for both the 2S (top) and 2P (bottom) sets.}
\label{fig:ionrates_vs_time}
\end{figure}

The measured photoionization intensities are plotted over time in Fig.~\ref{fig:ionrates_vs_time}. The experimental data was fitted using the solutions of Eq.~\ref{eq:ionization_rate}, incorporating the calculated values for $\Gamma_{i}$, $\Gamma_{i,j}$, and $R_{i}$. Additionally, an extra fitting term was introduced to account for other channels feeding into the $i^\mathrm{th}$ state that are not explicitly included in the model. Ideally, the contribution of this additional term should be minimal, indicating that all relevant states along the principal pathways have been considered. Indeed, for higher energy levels ($n>4$), the fits align well with our model, with only a minor contribution from the correction term. However, for the lowest energy level, the contribution from the additional fitting term is non-negligible. This is not surprising, as a significant number of feeding states might be omitted from the principal pathway.

\section{Conclusion}
In this study, we introduce an experimental method that enables three-dimensional high-resolution photoelectron spectroscopy where electrons are recorded over a large solid angle simultaneously. Unlike similar well-established techniques such as cold-target recoil ion momentum spectroscopy (COLTRIMS) or velocity map imaging (VMI), our approach not only allows for the determination of three-dimensional electron momentum vectors on an event-by-event basis without requiring a pulsed photon source, but it also provides access to the photoionization time with nanosecond resolution, thereby enabling to study time-dependent dynamics of atomic processes.

To demonstrate this capabilities, we studied the delayed ionization of $^{6}$Li atoms initially in the $2^{2}S_{1/2}$ and $2^{2}P_{3/2}$ states.
The atoms were excited into a polarized $F$ state by an ultrafast femtosecond laser pulse, from which they could decay to lower-lying states and/or eventually be ionized by absorbing a photon from the field of a continuous-wave optical dipole trap laser. The angular distributions of the photoelectrons ionized from the different populated states exhibit a significant shift relative to the laser polarization direction, an effect known as magnetic dichroism \cite{Acharya2021}. The measurement of the ionization time reveals the time dependence of the population dynamics, which is reasonably well reproduced by our model based on calculated electric-dipole transition rates. We attribute the remaining discrepancies between our measurements and calculations primarily to the omission of pathways and the limited knowledge of the amalgam of states populated by the femtosecond laser pulse.

The combination of femtosecond excitation and continuous-wave ionization, along with the capability to measure ionization times, constitutes a continuous pump-probe scheme. In conventional pump-probe experiments, the time dependence of the pumped system is typically probed by laboriously scanning the delay between two laser pulses. In contrast, the present method eliminates the need for such a scan, as the time dependence is directly obtained from the extracted ionization time. With its applicability to much longer timescales than femto- and attosecond pump-probe spectroscopy, this method serves as a valuable complement to existing time-resolved spectroscopic techniques.

While we focused on probing the incoherent population dynamics of the atomic system in this study, this technique also opens new avenues for exploring coherent processes that evolve on similar timescales. Unlike incoherent dynamics, which involve irreversible population redistribution, coherent phenomena such as Rydberg wave packets \cite{Lu2024}, Rabi dynamics \cite{Nandi2022}, spin-orbit wave packets \cite{Bayer2019}, and the superposition of Zeeman levels exhibit phase relationships making them particularly intriguing targets for future studies with our method. In this way, our approach not only extends the scope of time-resolved spectroscopy but can also help in developing coherent control schemes to precisely manipulate atomic dynamics.

\section*{Acknowledgments}
The experimental material presented here is based upon work supported by the National Science Foundation
under Grant \hbox{No.~PHY-1554776} and \hbox{No.~PHY-2207854}.

\bibliography{Rydberg_rev}

\begin{thebibliography}{29}%
\makeatletter
\providecommand \@ifxundefined [1]{%
 \@ifx{#1\undefined}
}%
\providecommand \@ifnum [1]{%
 \ifnum #1\expandafter \@firstoftwo
 \else \expandafter \@secondoftwo
 \fi
}%
\providecommand \@ifx [1]{%
 \ifx #1\expandafter \@firstoftwo
 \else \expandafter \@secondoftwo
 \fi
}%
\providecommand \natexlab [1]{#1}%
\providecommand \enquote  [1]{``#1''}%
\providecommand \bibnamefont  [1]{#1}%
\providecommand \bibfnamefont [1]{#1}%
\providecommand \citenamefont [1]{#1}%
\providecommand \href@noop [0]{\@secondoftwo}%
\providecommand \href [0]{\begingroup \@sanitize@url \@href}%
\providecommand \@href[1]{\@@startlink{#1}\@@href}%
\providecommand \@@href[1]{\endgroup#1\@@endlink}%
\providecommand \@sanitize@url [0]{\catcode `\\12\catcode `\$12\catcode `\&12\catcode `\#12\catcode `\^12\catcode `\_12\catcode `\%12\relax}%
\providecommand \@@startlink[1]{}%
\providecommand \@@endlink[0]{}%
\providecommand \url  [0]{\begingroup\@sanitize@url \@url }%
\providecommand \@url [1]{\endgroup\@href {#1}{\urlprefix }}%
\providecommand \urlprefix  [0]{URL }%
\providecommand \Eprint [0]{\href }%
\providecommand \doibase [0]{https://doi.org/}%
\providecommand \selectlanguage [0]{\@gobble}%
\providecommand \bibinfo  [0]{\@secondoftwo}%
\providecommand \bibfield  [0]{\@secondoftwo}%
\providecommand \translation [1]{[#1]}%
\providecommand \BibitemOpen [0]{}%
\providecommand \bibitemStop [0]{}%
\providecommand \bibitemNoStop [0]{.\EOS\space}%
\providecommand \EOS [0]{\spacefactor3000\relax}%
\providecommand \BibitemShut  [1]{\csname bibitem#1\endcsname}%
\let\auto@bib@innerbib\@empty
\bibitem [{\citenamefont {Manson}\ and\ \citenamefont {Starace}(1982)}]{Manson1982}%
  \BibitemOpen
  \bibfield  {author} {\bibinfo {author} {\bibfnamefont {S.~T.}\ \bibnamefont {Manson}}\ and\ \bibinfo {author} {\bibfnamefont {A.~F.}\ \bibnamefont {Starace}},\ }\href {https://doi.org/10.1103/revmodphys.54.389} {\bibfield  {journal} {\bibinfo  {journal} {Reviews of Modern Physics}\ }\textbf {\bibinfo {volume} {54}},\ \bibinfo {pages} {389} (\bibinfo {year} {1982})}\BibitemShut {NoStop}%
\bibitem [{\citenamefont {Reid}(2003)}]{Reid2003}%
  \BibitemOpen
  \bibfield  {author} {\bibinfo {author} {\bibfnamefont {K.~L.}\ \bibnamefont {Reid}},\ }\href {https://doi.org/10.1146/annurev.physchem.54.011002.103814} {\bibfield  {journal} {\bibinfo  {journal} {Annual Review of Physical Chemistry}\ }\textbf {\bibinfo {volume} {54}},\ \bibinfo {pages} {397} (\bibinfo {year} {2003})}\BibitemShut {NoStop}%
\bibitem [{\citenamefont {Bethe}\ and\ \citenamefont {Salpeter}(1957)}]{Bethe1957}%
  \BibitemOpen
  \bibfield  {author} {\bibinfo {author} {\bibfnamefont {H.~A.}\ \bibnamefont {Bethe}}\ and\ \bibinfo {author} {\bibfnamefont {E.~E.}\ \bibnamefont {Salpeter}},\ }\href {https://doi.org/10.1007/978-3-662-12869-5} {\emph {\bibinfo {title} {Quantum Mechanics of One- and Two-Electron Atoms}}}\ (\bibinfo  {publisher} {Springer Berlin Heidelberg},\ \bibinfo {year} {1957})\BibitemShut {NoStop}%
\bibitem [{\citenamefont {Liu}\ and\ \citenamefont {Jiang}(2024)}]{Liu2024}%
  \BibitemOpen
  \bibfield  {author} {\bibinfo {author} {\bibfnamefont {M.}~\bibnamefont {Liu}}\ and\ \bibinfo {author} {\bibfnamefont {W.-C.}\ \bibnamefont {Jiang}},\ }\href {https://doi.org/10.1364/oe.537644} {\bibfield  {journal} {\bibinfo  {journal} {Optics Express}\ }\textbf {\bibinfo {volume} {32}},\ \bibinfo {pages} {37703} (\bibinfo {year} {2024})}\BibitemShut {NoStop}%
\bibitem [{\citenamefont {Maroju}\ \emph {et~al.}(2023)\citenamefont {Maroju}, \citenamefont {Di~Fraia}, \citenamefont {Plekan}, \citenamefont {Bonanomi}, \citenamefont {Merzuk}, \citenamefont {Busto}, \citenamefont {Makos}, \citenamefont {Schmoll}, \citenamefont {Shah}, \citenamefont {Ribič}, \citenamefont {Giannessi}, \citenamefont {De~Ninno}, \citenamefont {Spezzani}, \citenamefont {Penco}, \citenamefont {Demidovich}, \citenamefont {Danailov}, \citenamefont {Coreno}, \citenamefont {Zangrando}, \citenamefont {Simoncig}, \citenamefont {Manfredda}, \citenamefont {Squibb}, \citenamefont {Feifel}, \citenamefont {Bengtsson}, \citenamefont {Simpson}, \citenamefont {Csizmadia}, \citenamefont {Dumergue}, \citenamefont {Kühn}, \citenamefont {Ueda}, \citenamefont {Li}, \citenamefont {Schafer}, \citenamefont {Frassetto}, \citenamefont {Poletto}, \citenamefont {Prince}, \citenamefont {Mauritsson}, \citenamefont {Callegari},\ and\ \citenamefont {Sansone}}]{Maroju2023}%
  \BibitemOpen
  \bibfield  {author} {\bibinfo {author} {\bibfnamefont {P.~K.}\ \bibnamefont {Maroju}}, \bibinfo {author} {\bibfnamefont {M.}~\bibnamefont {Di~Fraia}}, \bibinfo {author} {\bibfnamefont {O.}~\bibnamefont {Plekan}}, \bibinfo {author} {\bibfnamefont {M.}~\bibnamefont {Bonanomi}}, \bibinfo {author} {\bibfnamefont {B.}~\bibnamefont {Merzuk}}, \bibinfo {author} {\bibfnamefont {D.}~\bibnamefont {Busto}}, \bibinfo {author} {\bibfnamefont {I.}~\bibnamefont {Makos}}, \bibinfo {author} {\bibfnamefont {M.}~\bibnamefont {Schmoll}}, \bibinfo {author} {\bibfnamefont {R.}~\bibnamefont {Shah}}, \bibinfo {author} {\bibfnamefont {P.~R.}\ \bibnamefont {Ribič}}, \bibinfo {author} {\bibfnamefont {L.}~\bibnamefont {Giannessi}}, \bibinfo {author} {\bibfnamefont {G.}~\bibnamefont {De~Ninno}}, \bibinfo {author} {\bibfnamefont {C.}~\bibnamefont {Spezzani}}, \bibinfo {author} {\bibfnamefont {G.}~\bibnamefont {Penco}}, \bibinfo {author} {\bibfnamefont {A.}~\bibnamefont {Demidovich}}, \bibinfo {author} {\bibfnamefont {M.}~\bibnamefont {Danailov}}, \bibinfo {author} {\bibfnamefont {M.}~\bibnamefont {Coreno}}, \bibinfo {author} {\bibfnamefont {M.}~\bibnamefont {Zangrando}}, \bibinfo {author} {\bibfnamefont {A.}~\bibnamefont {Simoncig}}, \bibinfo {author} {\bibfnamefont {M.}~\bibnamefont {Manfredda}}, \bibinfo {author} {\bibfnamefont {R.~J.}\ \bibnamefont {Squibb}}, \bibinfo {author} {\bibfnamefont {R.}~\bibnamefont {Feifel}}, \bibinfo {author} {\bibfnamefont {S.}~\bibnamefont {Bengtsson}}, \bibinfo {author} {\bibfnamefont {E.~R.}\ \bibnamefont {Simpson}}, \bibinfo {author} {\bibfnamefont {T.}~\bibnamefont {Csizmadia}}, \bibinfo {author} {\bibfnamefont {M.}~\bibnamefont {Dumergue}}, \bibinfo {author} {\bibfnamefont {S.}~\bibnamefont {Kühn}}, \bibinfo {author} {\bibfnamefont {K.}~\bibnamefont {Ueda}}, \bibinfo {author} {\bibfnamefont {J.}~\bibnamefont {Li}}, \bibinfo {author} {\bibfnamefont {K.~J.}\ \bibnamefont {Schafer}}, \bibinfo {author} {\bibfnamefont {F.}~\bibnamefont {Frassetto}}, \bibinfo {author} {\bibfnamefont {L.}~\bibnamefont {Poletto}}, \bibinfo {author} {\bibfnamefont {K.~C.}\ \bibnamefont {Prince}}, \bibinfo {author} {\bibfnamefont {J.}~\bibnamefont {Mauritsson}}, \bibinfo {author} {\bibfnamefont {C.}~\bibnamefont {Callegari}},\ and\ \bibinfo {author} {\bibfnamefont {G.}~\bibnamefont {Sansone}},\ }\href {https://doi.org/10.1038/s41566-022-01127-3} {\bibfield  {journal} {\bibinfo  {journal} {Nature Photonics}\ }\textbf {\bibinfo {volume} {17}},\ \bibinfo {pages} {200} (\bibinfo {year} {2023})}\BibitemShut {NoStop}%
\bibitem [{\citenamefont {Vrakking}(2021)}]{Vrakking2021}%
  \BibitemOpen
  \bibfield  {author} {\bibinfo {author} {\bibfnamefont {M.~J.}\ \bibnamefont {Vrakking}},\ }\href {https://doi.org/10.1103/physrevlett.126.113203} {\bibfield  {journal} {\bibinfo  {journal} {Physical Review Letters}\ }\textbf {\bibinfo {volume} {126}},\ \bibinfo {pages} {113203} (\bibinfo {year} {2021})}\BibitemShut {NoStop}%
\bibitem [{\citenamefont {Silva}\ \emph {et~al.}(2021)\citenamefont {Silva}, \citenamefont {Atri-Schuller}, \citenamefont {Dubey}, \citenamefont {Acharya}, \citenamefont {Romans}, \citenamefont {Foster}, \citenamefont {Russ}, \citenamefont {Compton}, \citenamefont {Rischbieter}, \citenamefont {Douguet}, \citenamefont {Bartschat},\ and\ \citenamefont {Fischer}}]{Silva2021}%
  \BibitemOpen
  \bibfield  {author} {\bibinfo {author} {\bibfnamefont {A.~D.}\ \bibnamefont {Silva}}, \bibinfo {author} {\bibfnamefont {D.}~\bibnamefont {Atri-Schuller}}, \bibinfo {author} {\bibfnamefont {S.}~\bibnamefont {Dubey}}, \bibinfo {author} {\bibfnamefont {B.}~\bibnamefont {Acharya}}, \bibinfo {author} {\bibfnamefont {K.}~\bibnamefont {Romans}}, \bibinfo {author} {\bibfnamefont {K.}~\bibnamefont {Foster}}, \bibinfo {author} {\bibfnamefont {O.}~\bibnamefont {Russ}}, \bibinfo {author} {\bibfnamefont {K.}~\bibnamefont {Compton}}, \bibinfo {author} {\bibfnamefont {C.}~\bibnamefont {Rischbieter}}, \bibinfo {author} {\bibfnamefont {N.}~\bibnamefont {Douguet}}, \bibinfo {author} {\bibfnamefont {K.}~\bibnamefont {Bartschat}},\ and\ \bibinfo {author} {\bibfnamefont {D.}~\bibnamefont {Fischer}},\ }\href {https://doi.org/10.1103/physrevlett.126.023201} {\bibfield  {journal} {\bibinfo  {journal} {Physical Review Letters}\ }\textbf {\bibinfo {volume} {126}},\ \bibinfo {pages} {023201} (\bibinfo {year} {2021})}\BibitemShut {NoStop}%
\bibitem [{\citenamefont {Eickhoff}\ \emph {et~al.}(2021)\citenamefont {Eickhoff}, \citenamefont {Englert}, \citenamefont {Bayer},\ and\ \citenamefont {Wollenhaupt}}]{Eickhoff2021}%
  \BibitemOpen
  \bibfield  {author} {\bibinfo {author} {\bibfnamefont {K.}~\bibnamefont {Eickhoff}}, \bibinfo {author} {\bibfnamefont {L.}~\bibnamefont {Englert}}, \bibinfo {author} {\bibfnamefont {T.}~\bibnamefont {Bayer}},\ and\ \bibinfo {author} {\bibfnamefont {M.}~\bibnamefont {Wollenhaupt}},\ }\bibfield  {journal} {\bibinfo  {journal} {Frontiers in Physics}\ }\textbf {\bibinfo {volume} {9}},\ \href {https://doi.org/10.3389/fphy.2021.675258} {10.3389/fphy.2021.675258} (\bibinfo {year} {2021})\BibitemShut {NoStop}%
\bibitem [{\citenamefont {Pazourek}\ \emph {et~al.}(2015)\citenamefont {Pazourek}, \citenamefont {Nagele},\ and\ \citenamefont {Burgdörfer}}]{Pazourek2015}%
  \BibitemOpen
  \bibfield  {author} {\bibinfo {author} {\bibfnamefont {R.}~\bibnamefont {Pazourek}}, \bibinfo {author} {\bibfnamefont {S.}~\bibnamefont {Nagele}},\ and\ \bibinfo {author} {\bibfnamefont {J.}~\bibnamefont {Burgdörfer}},\ }\href {https://doi.org/10.1103/revmodphys.87.765} {\bibfield  {journal} {\bibinfo  {journal} {Reviews of Modern Physics}\ }\textbf {\bibinfo {volume} {87}},\ \bibinfo {pages} {765} (\bibinfo {year} {2015})}\BibitemShut {NoStop}%
\bibitem [{\citenamefont {Dörner}\ \emph {et~al.}(2000)\citenamefont {Dörner}, \citenamefont {Mergel}, \citenamefont {Jagutzki}, \citenamefont {Spielberger}, \citenamefont {Ullrich}, \citenamefont {Moshammer},\ and\ \citenamefont {Schmidt-Böcking}}]{Doerner2000}%
  \BibitemOpen
  \bibfield  {author} {\bibinfo {author} {\bibfnamefont {R.}~\bibnamefont {Dörner}}, \bibinfo {author} {\bibfnamefont {V.}~\bibnamefont {Mergel}}, \bibinfo {author} {\bibfnamefont {O.}~\bibnamefont {Jagutzki}}, \bibinfo {author} {\bibfnamefont {L.}~\bibnamefont {Spielberger}}, \bibinfo {author} {\bibfnamefont {J.}~\bibnamefont {Ullrich}}, \bibinfo {author} {\bibfnamefont {R.}~\bibnamefont {Moshammer}},\ and\ \bibinfo {author} {\bibfnamefont {H.}~\bibnamefont {Schmidt-Böcking}},\ }\href {https://doi.org/10.1016/s0370-1573(99)00109-x} {\bibfield  {journal} {\bibinfo  {journal} {Physics Reports}\ }\textbf {\bibinfo {volume} {330}},\ \bibinfo {pages} {95} (\bibinfo {year} {2000})}\BibitemShut {NoStop}%
\bibitem [{\citenamefont {Ullrich}\ \emph {et~al.}(2003)\citenamefont {Ullrich}, \citenamefont {Moshammer}, \citenamefont {Dorn}, \citenamefont {D{\"o}rner}, \citenamefont {Schmidt},\ and\ \citenamefont {Schmidt-B{\"o}cking}}]{Ullrich2003}%
  \BibitemOpen
  \bibfield  {author} {\bibinfo {author} {\bibfnamefont {J.}~\bibnamefont {Ullrich}}, \bibinfo {author} {\bibfnamefont {R.}~\bibnamefont {Moshammer}}, \bibinfo {author} {\bibfnamefont {A.}~\bibnamefont {Dorn}}, \bibinfo {author} {\bibfnamefont {R.}~\bibnamefont {D{\"o}rner}}, \bibinfo {author} {\bibfnamefont {L.~P.~H.}\ \bibnamefont {Schmidt}},\ and\ \bibinfo {author} {\bibfnamefont {H.}~\bibnamefont {Schmidt-B{\"o}cking}},\ }\href {https://doi.org/10.1088/0034-4885/66/9/203} {\bibfield  {journal} {\bibinfo  {journal} {Reports on Progress in Physics}\ }\textbf {\bibinfo {volume} {66}},\ \bibinfo {pages} {1463} (\bibinfo {year} {2003})}\BibitemShut {NoStop}%
\bibitem [{\citenamefont {Moshammer}\ \emph {et~al.}(2003)\citenamefont {Moshammer}, \citenamefont {Fischer},\ and\ \citenamefont {Kollmus}}]{Moshammer2003}%
  \BibitemOpen
  \bibfield  {author} {\bibinfo {author} {\bibfnamefont {R.}~\bibnamefont {Moshammer}}, \bibinfo {author} {\bibfnamefont {D.}~\bibnamefont {Fischer}},\ and\ \bibinfo {author} {\bibfnamefont {H.}~\bibnamefont {Kollmus}},\ }in\ \href {https://doi.org/10.1007/978-3-662-08492-2_2} {\emph {\bibinfo {booktitle} {Many-Particle Quantum Dynamics in Atomic and Molecular Fragmentation}}}\ (\bibinfo  {publisher} {Springer Berlin Heidelberg},\ \bibinfo {year} {2003})\ pp.\ \bibinfo {pages} {33--58}\BibitemShut {NoStop}%
\bibitem [{\citenamefont {Fischer}(2019)}]{Fischer2019}%
  \BibitemOpen
  \bibfield  {author} {\bibinfo {author} {\bibfnamefont {D.}~\bibnamefont {Fischer}},\ }in\ \href {https://doi.org/10.1515/9783110580297-006} {\emph {\bibinfo {booktitle} {Ion-Atom Collisions}}},\ \bibinfo {editor} {edited by\ \bibinfo {editor} {\bibfnamefont {M.}~\bibnamefont {Schulz}}}\ (\bibinfo  {publisher} {De Gruyter},\ \bibinfo {year} {2019})\ pp.\ \bibinfo {pages} {103--156}\BibitemShut {NoStop}%
\bibitem [{\citenamefont {Eppink}\ and\ \citenamefont {Parker}(1997)}]{Eppink1997}%
  \BibitemOpen
  \bibfield  {author} {\bibinfo {author} {\bibfnamefont {A.~T. J.~B.}\ \bibnamefont {Eppink}}\ and\ \bibinfo {author} {\bibfnamefont {D.~H.}\ \bibnamefont {Parker}},\ }\href {https://doi.org/10.1063/1.1148310} {\bibfield  {journal} {\bibinfo  {journal} {Review of Scientific Instruments}\ }\textbf {\bibinfo {volume} {68}},\ \bibinfo {pages} {3477} (\bibinfo {year} {1997})}\BibitemShut {NoStop}%
\bibitem [{\citenamefont {Dasch}(1992)}]{Dasch1992}%
  \BibitemOpen
  \bibfield  {author} {\bibinfo {author} {\bibfnamefont {C.~J.}\ \bibnamefont {Dasch}},\ }\href {https://doi.org/10.1364/ao.31.001146} {\bibfield  {journal} {\bibinfo  {journal} {Applied Optics}\ }\textbf {\bibinfo {volume} {31}},\ \bibinfo {pages} {1146} (\bibinfo {year} {1992})}\BibitemShut {NoStop}%
\bibitem [{\citenamefont {Vrakking}(2001)}]{Vrakking2001}%
  \BibitemOpen
  \bibfield  {author} {\bibinfo {author} {\bibfnamefont {M.~J.~J.}\ \bibnamefont {Vrakking}},\ }\href {https://doi.org/10.1063/1.1406923} {\bibfield  {journal} {\bibinfo  {journal} {Review of Scientific Instruments}\ }\textbf {\bibinfo {volume} {72}},\ \bibinfo {pages} {4084} (\bibinfo {year} {2001})}\BibitemShut {NoStop}%
\bibitem [{\citenamefont {Sharma}\ \emph {et~al.}(2018)\citenamefont {Sharma}, \citenamefont {Acharya}, \citenamefont {Silva}, \citenamefont {Parris}, \citenamefont {Ramsey}, \citenamefont {Romans}, \citenamefont {Dorn}, \citenamefont {de~Jesus},\ and\ \citenamefont {Fischer}}]{Sharma2018}%
  \BibitemOpen
  \bibfield  {author} {\bibinfo {author} {\bibfnamefont {S.}~\bibnamefont {Sharma}}, \bibinfo {author} {\bibfnamefont {B.~P.}\ \bibnamefont {Acharya}}, \bibinfo {author} {\bibfnamefont {A.~H. N. C.~D.}\ \bibnamefont {Silva}}, \bibinfo {author} {\bibfnamefont {N.~W.}\ \bibnamefont {Parris}}, \bibinfo {author} {\bibfnamefont {B.~J.}\ \bibnamefont {Ramsey}}, \bibinfo {author} {\bibfnamefont {K.~L.}\ \bibnamefont {Romans}}, \bibinfo {author} {\bibfnamefont {A.}~\bibnamefont {Dorn}}, \bibinfo {author} {\bibfnamefont {V.~L.~B.}\ \bibnamefont {de~Jesus}},\ and\ \bibinfo {author} {\bibfnamefont {D.}~\bibnamefont {Fischer}},\ }\href {https://doi.org/10.1103/physreva.97.043427} {\bibfield  {journal} {\bibinfo  {journal} {Physical Review A}\ }\textbf {\bibinfo {volume} {97}},\ \bibinfo {pages} {043427} (\bibinfo {year} {2018})}\BibitemShut {NoStop}%
\bibitem [{\citenamefont {Hubele}\ \emph {et~al.}(2015)\citenamefont {Hubele}, \citenamefont {Schuricke}, \citenamefont {Goullon}, \citenamefont {Lindenblatt}, \citenamefont {Ferreira}, \citenamefont {Laforge}, \citenamefont {Brühl}, \citenamefont {de~Jesus}, \citenamefont {Globig}, \citenamefont {Kelkar}, \citenamefont {Misra}, \citenamefont {Schneider}, \citenamefont {Schulz}, \citenamefont {Sell}, \citenamefont {Song}, \citenamefont {Wang}, \citenamefont {Zhang},\ and\ \citenamefont {Fischer}}]{Hubele2015}%
  \BibitemOpen
  \bibfield  {author} {\bibinfo {author} {\bibfnamefont {R.}~\bibnamefont {Hubele}}, \bibinfo {author} {\bibfnamefont {M.}~\bibnamefont {Schuricke}}, \bibinfo {author} {\bibfnamefont {J.}~\bibnamefont {Goullon}}, \bibinfo {author} {\bibfnamefont {H.}~\bibnamefont {Lindenblatt}}, \bibinfo {author} {\bibfnamefont {N.}~\bibnamefont {Ferreira}}, \bibinfo {author} {\bibfnamefont {A.}~\bibnamefont {Laforge}}, \bibinfo {author} {\bibfnamefont {E.}~\bibnamefont {Brühl}}, \bibinfo {author} {\bibfnamefont {V.~L.~B.}\ \bibnamefont {de~Jesus}}, \bibinfo {author} {\bibfnamefont {D.}~\bibnamefont {Globig}}, \bibinfo {author} {\bibfnamefont {A.}~\bibnamefont {Kelkar}}, \bibinfo {author} {\bibfnamefont {D.}~\bibnamefont {Misra}}, \bibinfo {author} {\bibfnamefont {K.}~\bibnamefont {Schneider}}, \bibinfo {author} {\bibfnamefont {M.}~\bibnamefont {Schulz}}, \bibinfo {author} {\bibfnamefont {M.}~\bibnamefont {Sell}}, \bibinfo {author} {\bibfnamefont {Z.}~\bibnamefont {Song}}, \bibinfo {author} {\bibfnamefont {X.}~\bibnamefont {Wang}}, \bibinfo {author} {\bibfnamefont {S.}~\bibnamefont {Zhang}},\ and\ \bibinfo {author} {\bibfnamefont {D.}~\bibnamefont {Fischer}},\ }\href {https://doi.org/10.1063/1.4914040} {\bibfield  {journal} {\bibinfo  {journal} {Review of Scientific Instruments}\ }\textbf {\bibinfo {volume} {86}},\ \bibinfo {pages} {033105} (\bibinfo {year} {2015})}\BibitemShut {NoStop}%
\bibitem [{\citenamefont {Metcalf}\ and\ \citenamefont {van~der Straten}(2007)}]{Metcalf2007}%
  \BibitemOpen
  \bibfield  {author} {\bibinfo {author} {\bibfnamefont {H.~J.}\ \bibnamefont {Metcalf}}\ and\ \bibinfo {author} {\bibfnamefont {P.}~\bibnamefont {van~der Straten}},\ }\href {https://doi.org/10.1002/9783527600441.oe005} {\bibinfo {title} {Laser cooling and trapping of neutral atoms}} (\bibinfo {year} {2007})\BibitemShut {NoStop}%
\bibitem [{\citenamefont {Silva}(2020)}]{Nish20}%
  \BibitemOpen
  \bibfield  {author} {\bibinfo {author} {\bibfnamefont {A.~H. N. C.~D.}\ \bibnamefont {Silva}},\ }\emph {\bibinfo {title} {Symmetry-breaking in the multi-photon ionization dynamics of oriented atoms}},\ \href {https://scholarsmine.mst.edu/doctoral_dissertations/2911} {Ph.D. thesis},\ \bibinfo  {school} {Missouri University of Science and Technology} (\bibinfo {year} {2020})\BibitemShut {NoStop}%
\bibitem [{\citenamefont {Grimm}\ \emph {et~al.}(2000)\citenamefont {Grimm}, \citenamefont {Weidemüller},\ and\ \citenamefont {Ovchinnikov}}]{Grimm2000}%
  \BibitemOpen
  \bibfield  {author} {\bibinfo {author} {\bibfnamefont {R.}~\bibnamefont {Grimm}}, \bibinfo {author} {\bibfnamefont {M.}~\bibnamefont {Weidemüller}},\ and\ \bibinfo {author} {\bibfnamefont {Y.~B.}\ \bibnamefont {Ovchinnikov}},\ }\bibinfo {title} {Optical dipole traps for neutral atoms},\ in\ \href {https://doi.org/10.1016/s1049-250x(08)60186-x} {\emph {\bibinfo {booktitle} {Advances In Atomic, Molecular, and Optical Physics}}}\ (\bibinfo  {publisher} {Elsevier},\ \bibinfo {year} {2000})\ pp.\ \bibinfo {pages} {95--170}\BibitemShut {NoStop}%
\bibitem [{\citenamefont {Wiley}\ and\ \citenamefont {McLaren}(1955)}]{Wiley1955}%
  \BibitemOpen
  \bibfield  {author} {\bibinfo {author} {\bibfnamefont {W.~C.}\ \bibnamefont {Wiley}}\ and\ \bibinfo {author} {\bibfnamefont {I.~H.}\ \bibnamefont {McLaren}},\ }\href {https://doi.org/10.1063/1.1715212} {\bibfield  {journal} {\bibinfo  {journal} {Review of Scientific Instruments}\ }\textbf {\bibinfo {volume} {26}},\ \bibinfo {pages} {1150} (\bibinfo {year} {1955})}\BibitemShut {NoStop}%
\bibitem [{\citenamefont {Acharya}\ \emph {et~al.}(2021)\citenamefont {Acharya}, \citenamefont {Dobson}, \citenamefont {Dubey}, \citenamefont {Romans}, \citenamefont {Silva}, \citenamefont {Foster}, \citenamefont {Russ}, \citenamefont {Bartschat}, \citenamefont {Douguet},\ and\ \citenamefont {Fischer}}]{Acharya2021}%
  \BibitemOpen
  \bibfield  {author} {\bibinfo {author} {\bibfnamefont {B.~P.}\ \bibnamefont {Acharya}}, \bibinfo {author} {\bibfnamefont {M.}~\bibnamefont {Dobson}}, \bibinfo {author} {\bibfnamefont {S.}~\bibnamefont {Dubey}}, \bibinfo {author} {\bibfnamefont {K.~L.}\ \bibnamefont {Romans}}, \bibinfo {author} {\bibfnamefont {A.~H. N. C.~D.}\ \bibnamefont {Silva}}, \bibinfo {author} {\bibfnamefont {K.}~\bibnamefont {Foster}}, \bibinfo {author} {\bibfnamefont {O.}~\bibnamefont {Russ}}, \bibinfo {author} {\bibfnamefont {K.}~\bibnamefont {Bartschat}}, \bibinfo {author} {\bibfnamefont {N.}~\bibnamefont {Douguet}},\ and\ \bibinfo {author} {\bibfnamefont {D.}~\bibnamefont {Fischer}},\ }\href {https://doi.org/10.1103/PhysRevA.104.053103} {\bibfield  {journal} {\bibinfo  {journal} {Phys. Rev. A}\ }\textbf {\bibinfo {volume} {104}},\ \bibinfo {pages} {053103} (\bibinfo {year} {2021})}\BibitemShut {NoStop}%
\bibitem [{\citenamefont {{\v{S}}ibali{\'{c}}}\ \emph {et~al.}(2017)\citenamefont {{\v{S}}ibali{\'{c}}}, \citenamefont {Pritchard}, \citenamefont {Adams},\ and\ \citenamefont {Weatherill}}]{Sibalic2017}%
  \BibitemOpen
  \bibfield  {author} {\bibinfo {author} {\bibfnamefont {N.}~\bibnamefont {{\v{S}}ibali{\'{c}}}}, \bibinfo {author} {\bibfnamefont {J.}~\bibnamefont {Pritchard}}, \bibinfo {author} {\bibfnamefont {C.}~\bibnamefont {Adams}},\ and\ \bibinfo {author} {\bibfnamefont {K.}~\bibnamefont {Weatherill}},\ }\href {https://doi.org/10.1016/j.cpc.2017.06.015} {\bibfield  {journal} {\bibinfo  {journal} {Computer Physics Communications}\ }\textbf {\bibinfo {volume} {220}},\ \bibinfo {pages} {319} (\bibinfo {year} {2017})}\BibitemShut {NoStop}%
\bibitem [{\citenamefont {Starace}(1982)}]{Starace1982}%
  \BibitemOpen
  \bibfield  {author} {\bibinfo {author} {\bibfnamefont {A.~F.}\ \bibnamefont {Starace}},\ }\href@noop {} {\emph {\bibinfo {title} {Encyclopedia of Physics Vol. 31, pp. 1-121}}},\ edited by\ \bibinfo {editor} {\bibfnamefont {W.}~\bibnamefont {Mehlhorn}}\ (\bibinfo  {publisher} {Springer-Verlag, Berlin},\ \bibinfo {year} {1982})\BibitemShut {NoStop}%
\bibitem [{\citenamefont {Marinescu}\ \emph {et~al.}(1994)\citenamefont {Marinescu}, \citenamefont {Sadeghpour},\ and\ \citenamefont {Dalgarno}}]{Marinescu1994}%
  \BibitemOpen
  \bibfield  {author} {\bibinfo {author} {\bibfnamefont {M.}~\bibnamefont {Marinescu}}, \bibinfo {author} {\bibfnamefont {H.~R.}\ \bibnamefont {Sadeghpour}},\ and\ \bibinfo {author} {\bibfnamefont {A.}~\bibnamefont {Dalgarno}},\ }\href {https://doi.org/10.1103/physreva.49.982} {\bibfield  {journal} {\bibinfo  {journal} {Physical Review A}\ }\textbf {\bibinfo {volume} {49}},\ \bibinfo {pages} {982} (\bibinfo {year} {1994})}\BibitemShut {NoStop}%
\bibitem [{\citenamefont {Lu}\ \emph {et~al.}(2024)\citenamefont {Lu}, \citenamefont {Wang}, \citenamefont {Kanungo}, \citenamefont {Yoshida}, \citenamefont {Dunning},\ and\ \citenamefont {Killian}}]{Lu2024}%
  \BibitemOpen
  \bibfield  {author} {\bibinfo {author} {\bibfnamefont {Y.}~\bibnamefont {Lu}}, \bibinfo {author} {\bibfnamefont {C.}~\bibnamefont {Wang}}, \bibinfo {author} {\bibfnamefont {S.~K.}\ \bibnamefont {Kanungo}}, \bibinfo {author} {\bibfnamefont {S.}~\bibnamefont {Yoshida}}, \bibinfo {author} {\bibfnamefont {F.~B.}\ \bibnamefont {Dunning}},\ and\ \bibinfo {author} {\bibfnamefont {T.~C.}\ \bibnamefont {Killian}},\ }\href {https://doi.org/10.1103/physreva.109.032801} {\bibfield  {journal} {\bibinfo  {journal} {Physical Review A}\ }\textbf {\bibinfo {volume} {109}},\ \bibinfo {pages} {032801} (\bibinfo {year} {2024})}\BibitemShut {NoStop}%
\bibitem [{\citenamefont {Nandi}\ \emph {et~al.}(2022)\citenamefont {Nandi}, \citenamefont {Olofsson}, \citenamefont {Bertolino}, \citenamefont {Carlström}, \citenamefont {Zapata}, \citenamefont {Busto}, \citenamefont {Callegari}, \citenamefont {Di~Fraia}, \citenamefont {Eng-Johnsson}, \citenamefont {Feifel}, \citenamefont {Gallician}, \citenamefont {Gisselbrecht}, \citenamefont {Maclot}, \citenamefont {Neoričić}, \citenamefont {Peschel}, \citenamefont {Plekan}, \citenamefont {Prince}, \citenamefont {Squibb}, \citenamefont {Zhong}, \citenamefont {Demekhin}, \citenamefont {Meyer}, \citenamefont {Miron}, \citenamefont {Badano}, \citenamefont {Danailov}, \citenamefont {Giannessi}, \citenamefont {Manfredda}, \citenamefont {Sottocorona}, \citenamefont {Zangrando},\ and\ \citenamefont {Dahlström}}]{Nandi2022}%
  \BibitemOpen
  \bibfield  {author} {\bibinfo {author} {\bibfnamefont {S.}~\bibnamefont {Nandi}}, \bibinfo {author} {\bibfnamefont {E.}~\bibnamefont {Olofsson}}, \bibinfo {author} {\bibfnamefont {M.}~\bibnamefont {Bertolino}}, \bibinfo {author} {\bibfnamefont {S.}~\bibnamefont {Carlström}}, \bibinfo {author} {\bibfnamefont {F.}~\bibnamefont {Zapata}}, \bibinfo {author} {\bibfnamefont {D.}~\bibnamefont {Busto}}, \bibinfo {author} {\bibfnamefont {C.}~\bibnamefont {Callegari}}, \bibinfo {author} {\bibfnamefont {M.}~\bibnamefont {Di~Fraia}}, \bibinfo {author} {\bibfnamefont {P.}~\bibnamefont {Eng-Johnsson}}, \bibinfo {author} {\bibfnamefont {R.}~\bibnamefont {Feifel}}, \bibinfo {author} {\bibfnamefont {G.}~\bibnamefont {Gallician}}, \bibinfo {author} {\bibfnamefont {M.}~\bibnamefont {Gisselbrecht}}, \bibinfo {author} {\bibfnamefont {S.}~\bibnamefont {Maclot}}, \bibinfo {author} {\bibfnamefont {L.}~\bibnamefont {Neoričić}}, \bibinfo {author} {\bibfnamefont {J.}~\bibnamefont {Peschel}}, \bibinfo {author} {\bibfnamefont {O.}~\bibnamefont {Plekan}}, \bibinfo {author} {\bibfnamefont {K.~C.}\ \bibnamefont {Prince}}, \bibinfo {author} {\bibfnamefont {R.~J.}\ \bibnamefont {Squibb}}, \bibinfo {author} {\bibfnamefont {S.}~\bibnamefont {Zhong}}, \bibinfo {author} {\bibfnamefont {P.~V.}\ \bibnamefont {Demekhin}}, \bibinfo {author} {\bibfnamefont {M.}~\bibnamefont {Meyer}}, \bibinfo {author} {\bibfnamefont {C.}~\bibnamefont {Miron}}, \bibinfo {author} {\bibfnamefont {L.}~\bibnamefont {Badano}}, \bibinfo {author} {\bibfnamefont {M.~B.}\ \bibnamefont {Danailov}}, \bibinfo {author} {\bibfnamefont {L.}~\bibnamefont {Giannessi}}, \bibinfo {author} {\bibfnamefont {M.}~\bibnamefont {Manfredda}}, \bibinfo {author} {\bibfnamefont {F.}~\bibnamefont {Sottocorona}}, \bibinfo {author} {\bibfnamefont {M.}~\bibnamefont {Zangrando}},\ and\ \bibinfo {author} {\bibfnamefont {J.~M.}\ \bibnamefont {Dahlström}},\ }\href {https://doi.org/10.1038/s41586-022-04948-y} {\bibfield  {journal} {\bibinfo  {journal} {Nature}\ }\textbf {\bibinfo {volume} {608}},\ \bibinfo {pages} {488} (\bibinfo {year} {2022})}\BibitemShut {NoStop}%
\bibitem [{\citenamefont {Bayer}\ \emph {et~al.}(2019)\citenamefont {Bayer}, \citenamefont {Gräfing}, \citenamefont {Kerbstadt}, \citenamefont {Pengel}, \citenamefont {Eickhoff}, \citenamefont {Englert},\ and\ \citenamefont {Wollenhaupt}}]{Bayer2019}%
  \BibitemOpen
  \bibfield  {author} {\bibinfo {author} {\bibfnamefont {T.}~\bibnamefont {Bayer}}, \bibinfo {author} {\bibfnamefont {D.}~\bibnamefont {Gräfing}}, \bibinfo {author} {\bibfnamefont {S.}~\bibnamefont {Kerbstadt}}, \bibinfo {author} {\bibfnamefont {D.}~\bibnamefont {Pengel}}, \bibinfo {author} {\bibfnamefont {K.}~\bibnamefont {Eickhoff}}, \bibinfo {author} {\bibfnamefont {L.}~\bibnamefont {Englert}},\ and\ \bibinfo {author} {\bibfnamefont {M.}~\bibnamefont {Wollenhaupt}},\ }\href {https://doi.org/10.1088/1367-2630/aafb87} {\bibfield  {journal} {\bibinfo  {journal} {New Journal of Physics}\ }\textbf {\bibinfo {volume} {21}},\ \bibinfo {pages} {033001} (\bibinfo {year} {2019})}\BibitemShut {NoStop}%
\end{thebibliography}%

\end{document}